\documentclass[preprint,aps,prd,showpacs,floatfix]{revtex4}
\usepackage{graphicx}
\usepackage{dcolumn}
\usepackage{bm}
\newcommand{\m}{mul\-ti\-pli\-ci\-ty }
\newcommand{\pbarp}{\mbox{$p\bar{p}$}}
\newcommand{\pt}{\mbox{$p_T$}}
\newcommand{\ptv}{\mbox{$p_T(V^{0})$}}
\newcommand{\ptm}{\mbox{$\langle p_T \rangle$}}
\newcommand{\ptvm}{\mbox{$\langle p_T(V^{0})\rangle$}}
\newcommand{\Et}{\mbox{$E_T$}}
\newcommand{\ecms}{\mbox{$E_{cms}$}}
\newcommand{\Nch}{\mbox{$N_{ch}^{\star}$}}
\newcommand{\lt}{\mbox{$\leq$}}
\newcommand{\gt}{\mbox{$\geq$}}
\newcommand{\ges}{\raisebox{-0.8ex}{$\sim$}\hspace{-1.8ex}\raisebox{0.2ex}{$>$}}
\newcommand{\les}{\raisebox{-0.8ex}{$\sim$}\hspace{-1.8ex}\raisebox{0.2ex}{$<$}}
\newcommand{\vv}{\mbox{$V^{0}$}}
\newcommand{\vvs}{\mbox{$V^{0}$'s}}
\newcommand{\kz}{\mbox{${\it K}^{0}_{s}$}}
\newcommand{\lz}{\mbox{$\Lambda^{0}$}}
\newcommand{\kl}{\mbox{${\it K}^{0}_{s} (\Lambda^{0})$}}
\newcommand{\gevc}{\mbox{GeV/{\it{c}}}}
\newcommand{\kdof}{\mbox{$\chi^{2}/N_{d.o.f.}$}}

\newcommand{\mpp}{\mbox{$\pm$}}
\begin{document}
\title{\kz\ and \lz\ Production Studies in  $p\bar{p}$ Collisions
at $\sqrt{s}$~=~1800 and 630 GeV }
\font\eightit=cmti8
\def\r#1{\ignorespaces $^{#1}$}
\author{
\hfilneg
\begin{sloppypar}
\noindent
D.~Acosta,\r {14} T.~Affolder,\r 7 M.G.~Albrow,\r {13} D.~Ambrose,\r {36}   
D.~Amidei,\r {27} K.~Anikeev,\r {26} J.~Antos,\r 1 
G.~Apollinari,\r {13} T.~Arisawa,\r {50} A.~Artikov,\r {11}
% ...
\end{sloppypar}
\vskip .026in
\begin{center}
\r 1  {\eightit Institute of Physics, Academia Sinica, Taipei, Taiwan 11529, 
Republic of China} \\
\r 2  {\eightit Argonne National Laboratory, Argonne, Illinois 60439} \\
\r 3  {\eightit Istituto Nazionale di Fisica Nucleare, University of Bologna,
I-40127 Bologna, Italy} \\
\r 4  {\eightit Brandeis University, Waltham, Massachusetts 02254} \\
\r 5  {\eightit University of California at Davis, Davis, California  95616} \\
\r 6  {\eightit University of California at Los Angeles, Los 
Angeles, California  90024} \\ 
\r 7  {\eightit University of California at Santa Barbara, Santa Barbara, California 
93106} \\ 
\r 8 {\eightit Instituto de Fisica de Cantabria, CSIC-University of Cantabria, 
39005 Santander, Spain} \\
\r 9  {\eightit Carnegie Mellon University, Pittsburgh, Pennsylvania  15213} \\
\r {10} {\eightit Enrico Fermi Institute, University of Chicago, Chicago, 
Illinois 60637} \\
\r {11}  {\eightit Joint Institute for Nuclear Research, RU-141980 Dubna, Russia}
\\
\r {12} {\eightit Duke University, Durham, North Carolina  27708} \\
\r {13} {\eightit Fermi National Accelerator Laboratory, Batavia, Illinois 
60510} \\
\r {14} {\eightit University of Florida, Gainesville, Florida  32611} \\
\r {15} {\eightit Laboratori Nazionali di Frascati, Istituto Nazionale di Fisica
               Nucleare, I-00044 Frascati, Italy} \\
\r {16} {\eightit University of Geneva, CH-1211 Geneva 4, Switzerland} \\
\r {17} {\eightit Glasgow University, Glasgow G12 8QQ, United Kingdom}\\
\r {18} {\eightit Harvard University, Cambridge, Massachusetts 02138} \\
\r {19} {\eightit Hiroshima University, Higashi-Hiroshima 724, Japan} \\
\r {20} {\eightit University of Illinois, Urbana, Illinois 61801} \\
\r {21} {\eightit The Johns Hopkins University, Baltimore, Maryland 21218} \\
\r {22} {\eightit Institut f\"{u}r Experimentelle Kernphysik, 
Universit\"{a}t Karlsruhe, 76128 Karlsruhe, Germany} \\
\r {23} {\eightit Center for High Energy Physics: Kyungpook National
University, Taegu 702-701; Seoul National University, Seoul 151-742; and
SungKyunKwan University, Suwon 440-746; Korea} \\
\r {24} {\eightit Ernest Orlando Lawrence Berkeley National Laboratory, 
Berkeley, California 94720} \\
\r {25} {\eightit University College London, London WC1E 6BT, United Kingdom} \\
\r {26} {\eightit Massachusetts Institute of Technology, Cambridge,
Massachusetts  02139} \\   
\r {27} {\eightit University of Michigan, Ann Arbor, Michigan 48109} \\
\r {28} {\eightit Michigan State University, East Lansing, Michigan  48824} \\
\r {29} {\eightit Institution for Theoretical and Experimental Physics, ITEP,
Moscow 117259, Russia} \\
\r {30} {\eightit University of New Mexico, Albuquerque, New Mexico 87131} \\
\r {31} {\eightit Northwestern University, Evanston, Illinois  60208} \\
\r {32} {\eightit The Ohio State University, Columbus, Ohio  43210} \\
\r {33} {\eightit Osaka City University, Osaka 588, Japan} \\
\r {34} {\eightit University of Oxford, Oxford OX1 3RH, United Kingdom} \\
\r {35} {\eightit Universita di Padova, Istituto Nazionale di Fisica 
          Nucleare, Sezione di Padova, I-35131 Padova, Italy} \\
\r {36} {\eightit University of Pennsylvania, Philadelphia, 
        Pennsylvania 19104} \\   
\r {37} {\eightit Istituto Nazionale di Fisica Nucleare, University and Scuola
               Normale Superiore of Pisa, I-56100 Pisa, Italy} \\
\r {38} {\eightit University of Pittsburgh, Pittsburgh, Pennsylvania 15260} \\
\r {39} {\eightit Purdue University, West Lafayette, Indiana 47907} \\
\r {40} {\eightit University of Rochester, Rochester, New York 14627} \\
\r {41} {\eightit Rockefeller University, New York, New York 10021} \\
\r {42} {\eightit Instituto Nazionale de Fisica Nucleare, Sezione di Roma,
University di Roma I, ``La Sapienza," I-00185 Roma, Italy}\\
\r {43} {\eightit Rutgers University, Piscataway, New Jersey 08855} \\
\r {44} {\eightit Texas A\&M University, College Station, Texas 77843} \\
\r {45} {\eightit Texas Tech University, Lubbock, Texas 79409} \\
\r {46} {\eightit Institute of Particle Physics, University of Toronto, Toronto
M5S 1A7, Canada} \\
\r {47} {\eightit Istituto Nazionale di Fisica Nucleare, University of Trieste/\
Udine, Italy} \\
\r {48} {\eightit University of Tsukuba, Tsukuba, Ibaraki 305, Japan} \\
\r {49} {\eightit Tufts University, Medford, Massachusetts 02155} \\
\r {50} {\eightit Waseda University, Tokyo 169, Japan} \\
\r {51} {\eightit University of Wisconsin, Madison, Wisconsin 53706} \\
\r {52} {\eightit Yale University, New Haven, Connecticut 06520} \\
\end{center}
}
\affiliation{CDF Collaboration}

\begin{abstract}
We present a study of the production of \kz\ and \lz\ in inelastic
\pbarp\ collisions at $\sqrt{s}$= 1800 and 630 GeV using data collected
by the CDF experiment at the Fermilab Tevatron.
Analyses of \kz\ and \lz\ multiplicity and transverse momentum
distributions, as well as of the dependencies of the average number 
and \ptm\ of \kz\ and \lz\ on charged particle multiplicity
are reported. Systematic comparisons are performed for the full sample
of inelastic collisions, and for the low and high momentum transfer
subsamples, at the two energies.
The \pt\ distributions extend above 8 \gevc, showing a \ptm\ 
higher than previous measurements.
The dependence of the mean \kl\ \pt\ on the charged particle multiplicity
for the three samples shows a behavior analogous
to that of charged primary tracks.
\end{abstract}
\maketitle
%
\section{INTRODUCTION}
\label{intro}

Hadron interactions are often classified as either ``hard" or
``soft"~\cite{ua2,sjo}. Although there is no formal definition for either,
the term ``hard interactions" 
denotes high momentum transfer parton-parton interactions
typically associated with such phenomena as jets of high energy transverse
to the incoming hadron momenta (\Et).
The ``soft'' interaction component encompasses everything else and
dominates the inelastic cross-section.
From a theoretical point of view, perturbative QCD provides a
reasonable description of high-\Et\ jet production.  However,
non-perturbative QCD, relevant to low-\Et\ hadronic production, is
not well understood.
Some QCD-inspired models~\cite{sjo} attempt to describe these processes 
by the superposition
of many parton interactions extrapolated to very low momentum transfers. 
It is not known, however, if these or other collective multi-parton processes 
are at work.
The experimental studies of low-\Et\ interactions are usually performed 
on data collected  using
minimum bias (MB) triggers, which, ideally, sample events in fixed proportion
to the production rate --- in other words, in their ``natural" distribution.
Lacking a comprehensive description of the microscopic processes~\cite{mod}
involved in
low-\Et\ interactions,  our knowledge of the details of low transverse
momentum (\pt) particle production rests largely upon empirical connections
between phenomenological models  
and data collected with MB triggers at many center-of-mass energies (\ecms). 
Such comparisons necessarily face the difficulty of isolating events
of a purely ``soft" or ``hard" nature.

Comparative studies of the event structure through collective
variables such as the charged particle multiplicity
and the transverse energy of the event are important to our
understanding of the soft production  mechanism.
In a previous paper~\cite{noi},
a novel approach in addressing this issue using samples of
\pbarp\ collisions at $\sqrt{s} = 1800$ and 630 GeV collected with an MB
trigger was described. The analysis divided the full MB 
samples into two
subsamples, one highly enriched in soft interactions, the other in hard
interactions. Comparisons between the subsamples and as a function of
\ecms\ were performed. 
The same approach has been applied here to the production of strange 
particles.

Beside gluons and the lighter quarks {\it u} and {\it d}, strange quark 
production is the only component of low-\pt\ 
multiparticle interactions which is statistically significant and 
experimentally accessible with an MB trigger. 
It is also a probe for investigating the transition of soft hadron 
interactions to the QCD high-\pt\ perturbative region.

This paper describes a study of \kz\ and \lz\ production in \pbarp\
interactions at different \ecms.
Inclusive distributions of the 
multiplicity and transverse momentum of \kz\ and \lz\ are presented first.
The high statistics of the data sample collected 
at $\sqrt{s}$= 1800 and 630 GeV allow an extension of the range and 
precision of these measurements with respect to previous ones.
Studies of the dependence of the average \pt\ of \kl\ and of their mean number
on the event charged multiplicity are also presented.
Different behavior of the hard and soft subsamples is observed, 
consistent with prior reports on charged particles~\cite{noi}.
 
\section{DATA COLLECTION}
\label{evsel}
\subsection{The CDF Detector}
\label{cdfdet}

Data samples have been collected with the CDF detector at the Fermilab 
Tevatron Collider. The CDF apparatus has been described 
elsewhere~\cite{detec};
here only the parts of the detector utilized for the present analysis 
are discussed.
The coordinate system is defined with respect to the proton beam direction,
which defines the positive $z$ direction, while the azimuthal angle $\phi$
is measured around the beam axis. The polar angle $\theta$ is measured
with respect to the positive $z$ direction. The pseudorapidity, $\eta$, is
often used and is defined as $\eta = -\ln(\tan[\theta/2])$. 
Transverse components of particle energy and momentum are conventionally
defined as projections onto the plane transverse to the beam line, 
$E_T=E\sin{\theta}$ and $p_T= |\vec{p}|\sin{\theta}$.

Data were collected with an MB trigger at 1800 GeV during Runs 1A (1992-93) 
and 1B (1994-95), and at 1800 and 630 GeV during Run 
1C (1995-96).
This trigger requires coincident hits in scintillator counters,
located at 5.8 m from either side down stream of the nominal interaction 
point and covering the pseudorapidity
interval $3.2 < |\eta| < 5.9$, in coincidence with a beam-crossing.
\par
The analysis uses charged tracks reconstructed within the Central Tracking
Chamber (CTC).
The CTC is a cylindrical drift chamber 
covering an $\eta$ interval of about three units with high efficiency for
$|\eta|\leq 1$ and $p_T \geq$ 0.4 \gevc.
The inner radius of the CTC is 31.0 cm and the outer radius is 
132.5 cm. The full CTC volume is contained in a superconducting solenoidal
magnet which operates at 1.4 T~\cite{magnet}.
The CTC has 84 sampling wire layers, organized in 5 axial and 4 stereo 
``superlayers"~\cite{ctc}. 
Axial superlayers have 12 radially  separated layers of sense wires,
parallel to the $z$-axis (the beam axis), that measure the  $r$-$\phi$
position of a track. Stereo superlayers have 6 sense wire layers, with a
$\sim$ 3$^{\circ}$ stereo angle, which measure a combination of $r$-$\phi$
and  $z$ positions. The stereo angle direction alternates with each 
neighboring stereo superlayer. 
Measurements from axial and stereo superlayers are combined to form a
 three-dimensional track.
The spatial resolution of each point measurement in the CTC is less than 
200 $\mu$m;
the transverse momentum resolution, including multiple-scattering effects, 
is $\sigma_{p_{T}}$/$p_{T}^{2}\leq 0.003$ $(\gevc)^{-1}$.
\par 
Inside the CTC inner radius, a set of time-projection chambers 
(VTX)~\cite{vtx}
provides $r$-$z$ tracking information out to a
radius of 22 cm for $|\eta| < 3.25$. The VTX is used in this analysis to 
find the $z$ positions of event vertices, defined as sets of tracks
converging to the same point along the $z$-axis.
The closest detector to the beam-pipe is the Silicon Vertex Finder (SVX),
used to reconstruct vertex positions in the transverse view.
Reconstructed vertices are classified as
either ``primary" or ``secondary" based upon several parameters:
a minimum of 4 converging track segments in $|\eta|<3$ (a track segment is
a sequence of 4 aligned hits),
the total number of hits used to form a segment, forward-backward
symmetry and vertex isolation.
Isolated, higher multiplicity vertices with highly symmetric
topologies are typically classified as primary; lower multiplicity,
highly asymmetric vertices, or those with few hits in the reconstructed
tracks, are typically classified as secondary.
Systematic uncertainties introduced by the vertex classification scheme are
discussed in Section~\ref{syserr}.

The transverse energy flux is measured by a
calorimeter system~\cite{calor} covering $|\eta| \leq$ 4.2.
The calorimeter consists of three sub-systems, each with separate 
electromagnetic and hadronic components:
the central calorimeter, covering the range
$|\eta|<$1.1; the end-plug, covering 1.1$<|\eta|<$2.4; and the
forward calorimeter, covering 2.2$<|\eta|<$4.2. Energy measurements are
made within projective ``towers" that span 0.1 units of $\eta$ and
15$^{\circ}$(5$^{\circ}$) in $\phi$
within the central (end-plug and forward) calorimeter.

\subsection{The Data Set}
\label{dataset}
The 1800 GeV MB data sample consists of subsamples  collected
during three different time periods.  Approximately 1.7$\times10^{6}$ 
events were collected in Run  1A at an average luminosity of 
3.3$\times10^{30}$ s$^{-1}$cm$^{-2}$, 1.5$\times10^{6}$  in Run 1B 
at an average luminosity of 9.1$\times10^{30}$ s$^{-1}$cm$^{-2}$  
and 1.06$\times10^{5}$ in Run 1C at an average
luminosity of 9.0$\times10^{30}$  s$^{-1}$cm$^{-2}$. 
The 630 GeV data set consists of about 2.6$\times10^{6}$ events recorded
during Run 1C at an average
luminosity of 1.3$\times10^{30}$ s$^{-1}$cm$^{-2}$.   

Additional event selection conducted offline removed the following
events:
(i) events identified as containing cosmic ray particles as determined by
   time-of-flight measurements using
   scintillator counters in the  central calorimeter;
(ii) events with no reconstructed tracks;
(iii) events exhibiting symptoms of known calorimeter problems;
(iv) events with at least one charged particle reconstructed in the CTC
   to have \pt\ \gt\ 400 MeV/{\it c}, but no central calorimeter tower with 
   energy deposition above 100 MeV;
(v) events with more than one primary vertex;
(vi) events with a primary vertex more than 60 cm away from the center of the
    detector (in order to ensure uniform acceptance in the assumed fiducial
    region and good track and calorimeter energy reconstruction); and 
(vii) events with no primary vertices.

After all event selection requirements, 2,079,558 events remain in the full
MB sample at $\sqrt{s} = 1800$ GeV  and 1,963,157 at $\sqrt{s} = 630$ GeV. 
The vast majority of rejected events failed the vertex selection. About 0.01\%
of selected events contain background tracks from cosmic rays that are
coincident in time with the beam crossing and pass near the event vertex. The
residual contamination due to the interactions of the beam particles  with
the gas in the beam pipe is about 0.02\%.
A more detailed discussion of the systematic uncertainties
arising from the event selection criteria and other sources is presented in
Ref.~\cite{noi}.

\section{CHARGED TRACKS and \kl\ SELECTION}
\label{k0sel}

We require all reconstructed tracks to pass through a minimum number of layers
in the CTC and have a minimum number of hits in each superlayer in order
to reduce the number of misreconstructed tracks and those with large 
reconstruction uncertainties. 
The remaining track set, which includes primary and secondary tracks,
is used as a starting point for both the selection of primary charged
tracks and for the \kl\ candidate identification procedure.

\vspace{1ex}
{\sc Charged track multiplicity definition.} 
Tracks are required to pass within 0.5 cm of the beam axis, and
within 5~cm along the $z$-axis from the primary event vertex.
In order to ensure high efficiency and acceptance, tracks are accepted only
if they satisfy the conditions \pt\gt~0.4~\gevc\ and
$|\eta|$ \lt~1.0 .
This selection defines the charged track multiplicity in an event, \Nch.

\vspace{1ex}
{\sc \kz\ and \lz\ selection.}
\kz\ and \lz~\cite{llbar}(from now on collectively referred to
as \vv) are selected
looking for opposite-charge pairs of tracks converging to a common vertex
displaced from the beam line in the transverse direction.
A vertex fit is performed to ensure that the two tracks originate from the 
same vertex. A candidate is required to have a fit probability greater 
than 5\%. In a further step
a fit is performed constraining the \vv\ momentum vector (within the track
uncertainties) to point in the direction of the primary vertex (pointing
constraint fit). The candidates are kept if the fit probability is greater
than 5\% and the recomputed invariant mass is within three standard deviations
of the word average \kz\ or \lz\ mass~\cite{pdg}.

The analysis selection also requires:
\begin{itemize}
\item $L_{xy}$(\vv)$\geq$1~cm, where  $L_{xy}$ is the distance from primary
         vertex to the decay vertex of the
        \vv\ in the $r$-$\phi$ plane;
\item both decay tracks have $|\eta|\leq$~1.5 and $p_T\geq$~0.3~\gevc;
\item the \vv\ line-of-flight is close to the event vertex along the $z$ 
         axis:
         $|z_{0}^{V^{0}} - z_{0}^{vertex}| <$6~cm;
\item impact parameter $d_{0}$(\vv)~$<$~0.7~cm;
\item $p_{T}(V^0)\geq$0.4~\gevc\ and $|\eta(V^0)|\leq$1.0
\end{itemize}

For events with more than one \vv\  candidate sharing the same track,
only the candidate with the lower vertex fit \kdof\ is retained.

After all selection requirements, we find 36,642 \kz\ and 7,518 \lz\ in the 
1800 GeV MB sample and 32,222 \kz\ and 5,883 \lz\ in the 630 GeV MB sample 
(see Table~\ref{tab1}).

\begin{figure}
\scalebox{1}{
\includegraphics[0,24][390,290]{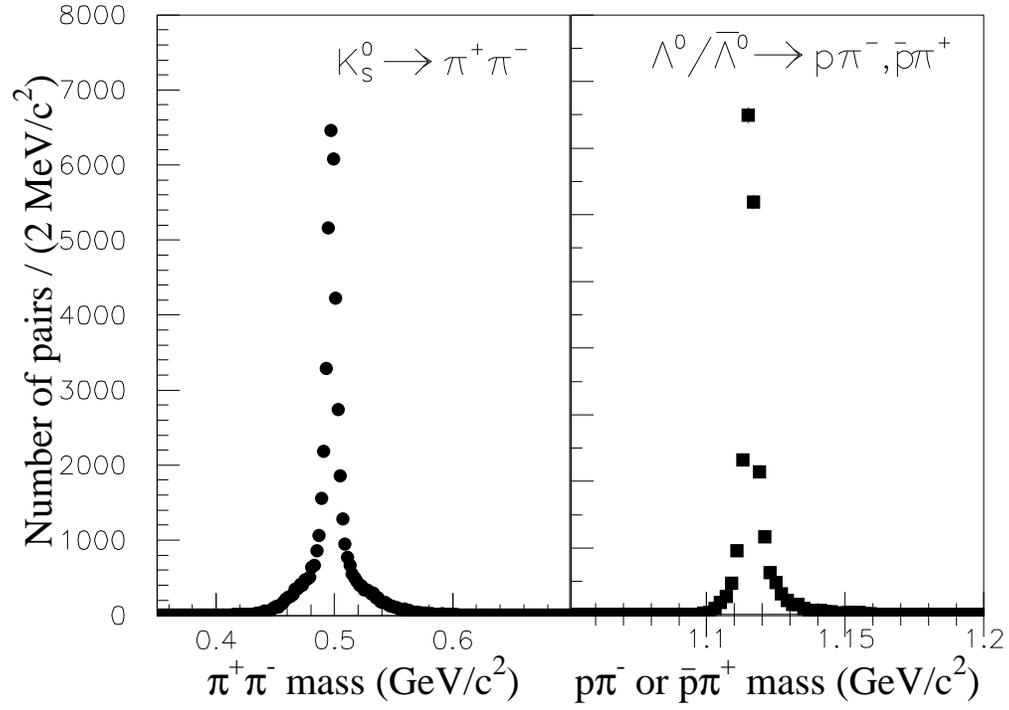}}
\caption{\label{fig:klmass}Invariant mass with a $\pi-\pi$ and $\pi$-p mass 
         assignment for oppositely signed track pairs passing all selection
         requirements but allowing the mass to be within 10$\sigma$ from the 
         nominal mass. 
	 The background includes the contamination by \kz\ in the
         \lz\ sample and vice versa. 1800 GeV data are shown.}
\end{figure}

The invariant mass distributions of the \kz\ and \lz\ surviving the 
selection requirements, but with the mass window extended to ten standard 
deviations from the world average, are shown in Fig.~\ref{fig:klmass}; 
in both cases the peaks are narrow but, because of the fit procedure,
the background is not flat and may not be accounted for by the
level of the sidebands. 
We also note that this background includes
the contamination of \kz\ in the \lz\ sample and vice versa.
A detailed background evaluation is discussed in Section~\ref{effcorr}. 

\section{SELECTION OF {\it SOFT} AND {\it HARD} INTERACTIONS }
\label{shsel}

The identification of ``soft" and ``hard" interactions
is largely a matter of
definition~\cite{multiparton} since it is unknown how to distinguish soft 
and hard parton interactions. This is true
from both the theoretical and experimental points of view.
In this analysis, we use a jet
reconstruction algorithm to define the two cases. The
algorithm employs a cone with radius  
$ R=(\Delta\eta^2 + \Delta\phi^2)^{1/2}=0.7$ 
to define ``clusters" of calorimeter
towers belonging to a jet. 
To be considered, a cluster must have a transverse energy \Et, defined as
the scalar sum of the transverse energy of all the towers included in
the cone, of at least 1 GeV in a seed tower, plus at least 0.1 GeV in an 
adjacent tower.

In the regions $|\eta| < 0.02$ and $1.1 < |\eta| < 1.2$, a track-clustering
algorithm is used instead of the calorimeter algorithm to compensate
for energy lost in calorimeter cracks.
A track cluster is defined as one track with $p_T>$ 0.7 \gevc\ and at least
one other track with $p_T \geq$0.4 \gevc\ in a cone of radius $R = 0.7$.

We define a {\it soft} event as one that contains no cluster with
$E_T > 1.1$ GeV. All other events are classified as {\it hard}.

\section{EFFICIENCY and CORRECTIONS}
\label{effcorr}

The probability of observing a real \vv\ in the apparatus is influenced 
by several effects. In this section we discuss the efficiency of
track reconstruction, the correction for limited
acceptance, and evaluation of the background.
At the end, some cross-checks of
the correction procedures are also briefly described.
 
1. The efficiency for finding \kl\ has been investigated in two 
different ways. 
In the first method, simulated hits from singly-generated \vv\ 
are embedded among the set of hits of MB events from the data. 
The events are then reconstructed with standard \vv\ search and selection.
In the second method, entire MB events with \vv\ production and decay are
generated with  {\sc pythia}/{\sc jetset} Monte Carlo
(MC)~\cite{noi},~\cite{pythia}. 
Full CDF detector simulation and reconstruction are then applied to the 
events and the resulting reconstructed kinematic distributions are 
similar to those observed in the data.
The results from the two methods are compatible 
within the statistical uncertainties.

The efficiency is defined as  the ratio of the reconstructed to 
the generated number of \vv\ in the fiducial region. 
It is examined as a function of single kinematical variables 
of the \vv, integrating over all the remaining variables.
The embedding method, given its almost flat \vv\ distribution in all
variables, gives smooth and statistically better determined efficiency 
dependences from all observables over all the acceptance limits.
The results of this method are used to determine the shape of the
efficiency as a function of any chosen variable. 
Each efficiency distribution from the embedding method is then scaled by
an overall normalization factor so that the integrated efficiency 
obtained from the embedding method matches the integrated efficiency 
from the full MC method. 

The efficiency for finding a single \kl\ is approximately constant (around 40\%
(32\%)) as a function of $\eta$(\vv) in the region of 
$|\eta(V^{0})|<$1 and \pt(\vv)$>$0.4 \gevc. 
As a function of \ptv\ (in the same $\eta$ region),
the efficiency rises rapidly from 25\%  (15\%) at 
0.4 \gevc\ to about 50\% (40\%) for \pt$\sim$1 \gevc,
and then slowly decreases to $\simeq$~20\% ($\simeq$~15\%) 
for \pt\ \ges\ 8 \gevc.
This behavior is due to the difficulty in reconstructing low-\pt\ secondary
tracks and in identifying secondary vertices far from the primary
vertex.
The efficiency also diminishes for $L_{xy}$ \les~3~cm, 
while it is roughly constant 
as a function of the charged multiplicity of the event.
The overall efficiency is about 39\% for \kz\ and 31\% for \lz.

2. A correction for the fiducial acceptance requirement in 
$L_{xy}$ and in the
\pt\ of the \kl\ decay products is estimated using MC and found to
range from about 15(20) at \pt=~0.5(1)~\gevc\ to about 1
for \pt\ \ges\ 5 \gevc.

3. The contamination by \lz\ in the \kz\ sample is estimated to be 
$\simeq$~3\% as found in the {\sc pythia} MC simulation; 
the contamination by \kz\ in the \lz\ sample is about 7\% on average while
it is almost 50\% for \pt(\lz) $<$ 1.5 \gevc.
The same MC sample is used to compute 
the probability of selecting fake secondary vertices (not due to \kz\ or \lz\ 
decays). Such probability is found to account for roughly 25\% of the 
\kz\ and 40\% of the \lz.

\begin{figure}{
\scalebox{1}{
\includegraphics[140,260][426,554]{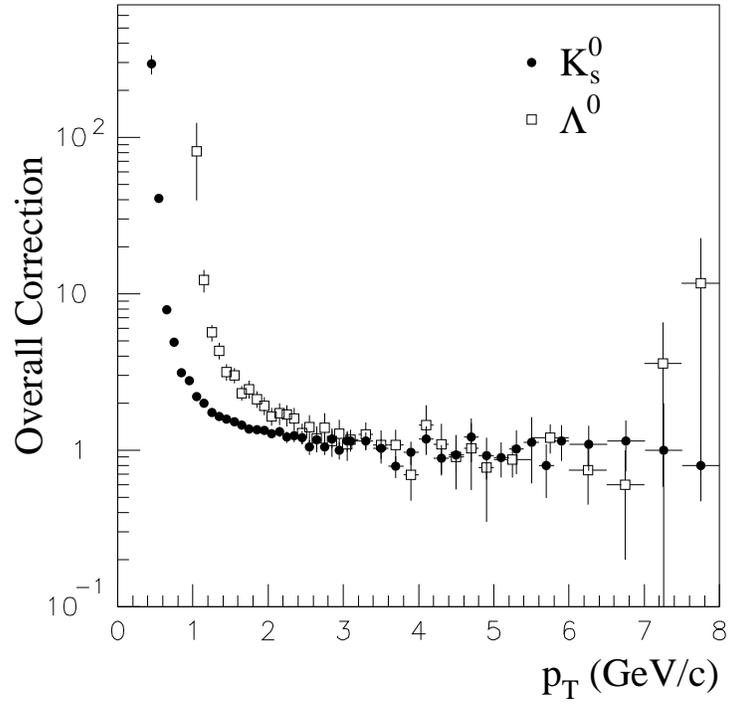}} }
\caption{\label{fig:correction}Overall correction factors at 1800 GeV for \kl\
 as a function of the transverse momentum. The correction factors are defined
 in the text.} 
\end{figure}

4. The overall correction factor for a generic inclusive variable 
X (e.g. the \ptv) is given by the expression:
\begin{equation}
 C(X) = \frac{1 - R_{fake}(X)}{\epsilon (X) \times A(X)} 
\label{eq:fpt1}
\end{equation}
\noindent
where $R_{fake}$ is the probability of a fake \vv, $\epsilon$ is the global 
efficiency and $A$ is the acceptance. The overall correction factors, 
as a function of \ptv, are shown in Fig.~\ref{fig:correction}. 
The integrated MC correction factors are estimated to
be 4.5$\pm$0.1 and 10.1$\pm$0.2 for \kz\ and \lz\ respectively.

5. Because of the small differences that exist between some {\sc pythia}
distributions and the data, we expect that the MC correction will not be 
fully reliable in the regions where it changes very rapidly. 
Evidence of this is given by 
the reconstructed \vv\ \pt\ versus the proper time which  
shows a depletion in the low-\pt\ and low-lifetime region, even after 
applying the MC correction.

We use the following method to correct the counted number of \kl\ in this
region.
The \kl\ invariant \pt\ distributions for the full MB sample are fitted 
with a functional power-law form:
\begin{equation}
 E \frac{d^{3}N_{V^0}}{dp^{3}} = A \left( \frac{p_{0}}{p_{0}+\pt}\right) ^{n}, 
\label{eq:fpt2}
\end{equation}
\noindent
where E is the particle energy and $p_{0}$, $A$ and $n$ are free parameters, 
in the region above 0.8 \gevc\ (1.1 \gevc\ for \lz).
This equation has been widely used to fit the \pt\ distributions of charged
tracks down to the lower measured \pt\ \cite{ptfit}. 
The fitted function is extrapolated down to \ptv=0.4 \gevc\ and
the corrected number of \kl\ is extracted from the integral of the
curve. \par\noindent
In the full MB sample, the number of undetected \vv\ is estimated
to be approximately $18\times10^{3}$ \kz\ and $14\times10^{3}$ \lz\ at 
1800 GeV, and $24\times10^{3}$ and $12\times10^{3}$, 
respectively, at 630 GeV.

6. The above correction affects the measurements of the mean number
of \vv\ per event and of the mean \pt\ when computed at fixed \Nch. 
The latter is 
calculated as the sum of the \pt's of \kl, above 0.4~\gevc,  
observed in events of a given charged multiplicity, divided by 
the number of \kl:
\begin{equation}
\langle p_{T}\rangle=\frac{1}{N_{V^0}}\sum_i^{N_{V^0}} p_{{T}_{i}}
\label{eq:fpt3}
\end{equation}
 
An estimate of the number of undetected \vv\ and the resulting effect on
the \ptvm\ and on the \vv\ multiplicity are obtained, for each \ptv\ 
distribution, with the procedure used in the inclusive case.
The constraint that the sum of undetected \vv\ for each multiplicity should 
give the number of undetected \vv\ computed from the inclusive \pt\ 
distribution is imposed.  
The corrected \ptvm\ is computed by extrapolating the fitted \ptv\
distribution down to \ptv=0.4 \gevc.

7. The consistency of the correction procedures described above has been
verified through the following cross-checks.

\begin{figure}{
\scalebox{1}{
\includegraphics[2,18][400,294]{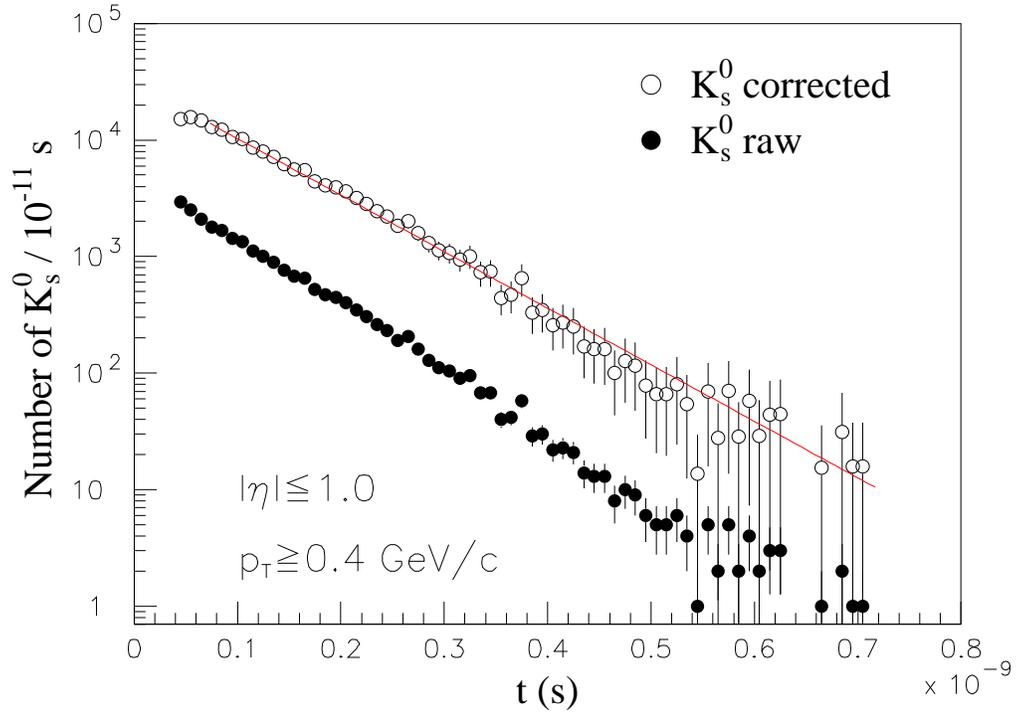}}
\vspace{-2ex}
\scalebox{1}{
\includegraphics[2,18][400,294]{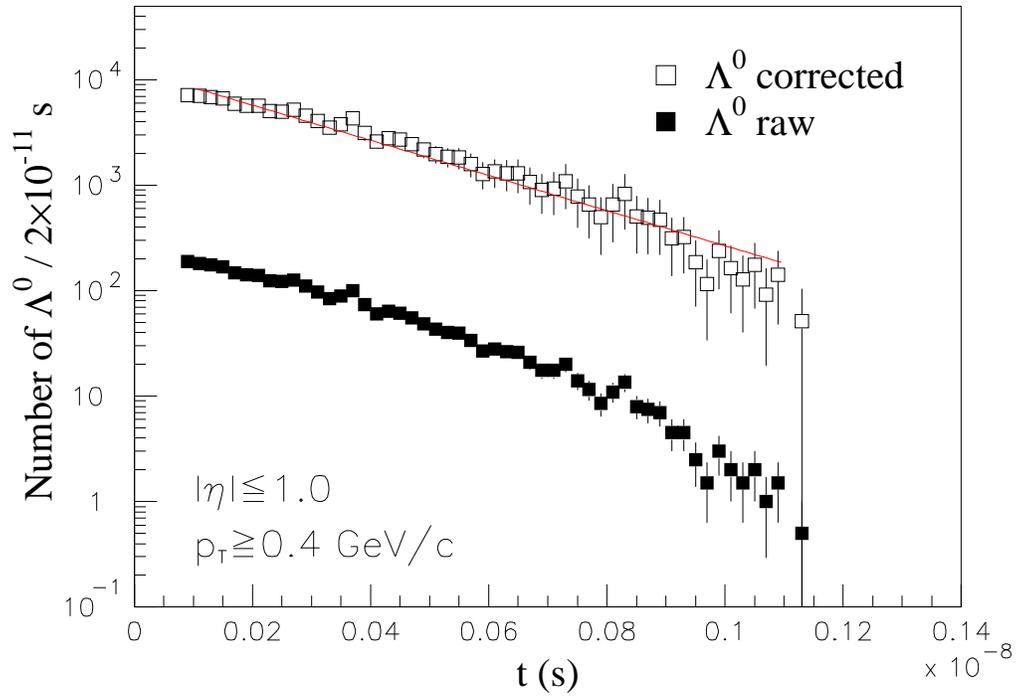}} }
\caption{\label{fig:lifetimes}Raw and corrected \kz\ and \lz\ proper lifetime 
  distributions for 1800 GeV data. The line represents the best exponential 
  fit to the data.}
\end{figure}

In order to check the selection requirements and the quality of the 
efficiency 
correction, the raw and corrected proper lifetime distributions at 1800 GeV
are shown in Fig.~\ref{fig:lifetimes}.
Fitting to an exponential form gives a \kz\ mean proper lifetime of 
(0.89 $\pm 0.01)\times10^{-10}$ s (\kdof=49.7/59) for 1800 GeV
and (0.90 $\pm 0.01)\times10^{-10}$ s (\kdof=60.6/56) for 630 GeV.
Both values are consistent with the world average values~\cite{pdg}.
The same fit to the \lz\ proper lifetime distributions gives a mean of
(2.61 $\pm 0.07)\times10^{-10}$ s (\kdof=44.2/49) for 1800 GeV and 
(2.61 $\pm 0.07)\times10^{-10}$ s (\kdof=57.4/50) for 630 GeV. 
The proper lifetime regions used for the fit are $\tau>0.7\times10^{-10}$ s 
(\kz) and $\tau>10^{-10}$ s (\lz).

The number of undetected \vv\ extracted from the fitted \pt\ curve is 
also checked. The proper lifetime distributions 
of Fig.~\ref{fig:lifetimes} are fitted to an exponential form with fixed 
slope (the \kz/\lz\ mean lifetimes~\cite{pdg}, 
$\tau_{K^{0}_{s}}=0.8935\times 10^{-10}$ s; 
 $\tau_{\Lambda^{0}}=2.632\times 10^{-10}$ s) 
in the region 
$\tau > 0.7\times10^{-10}$ s ($\tau >10^{-10}$ s for \lz) 
and the fitted curves are integrated down to $\tau=0$. 
The number of undetected \kl\ obtained matches to 15\% (30\%) with the number 
from the \pt\ distribution.
Furthermore 
the \pt\ distributions of \kl\ with proper lifetimes greater than 
$0.8\times10^{-10} s$ ($1.0\times10^{-10} s$ for \lz) are compared 
with the corresponding  distributions for all lifetimes;
the comparison gives the same values of average \pt. 
When normalized to one another, the curves give a 
comparable number of \kl\ in the extrapolated region.

An additional cross-check for correcting the average \pt\ of the \vv\ 
observed in events of fixed multiplicity consists of plotting
the proper lifetime distribution in slices of \pt\ so that each distribution 
corresponds to one bin in \pt. This is done for each bin in multiplicity.
After fitting the distribution in the long lifetime region in each
\pt\ bin, the correct number of \kz\ in the short lifetime region 
can be extrapolated from the fits.
The \ptvm\ can then
be recomputed from the modified \pt\ distribution. 
The \pt\ values obtained using the two different correction methods are 
consistent.
In the case of \lz, no events are found with \pt\ below 1 \gevc\ due 
to the tight fiducial requirements imposed in the analysis. Therefore, in
the \lz\ case, the correction method based on extrapolating the proper 
lifetime distribution at each \pt\ bin cannot be used.
Because of this, the cross-checks are limited to comparing 
the number of extrapolated  \lz\ in the \pt\ and proper lifetime distributions
of the full data sample.

\vspace{1ex}
Finally, we refer to~\cite{noi} for a detailed discussion of the charged 
track selection and reconstruction efficiencies.

\section{SYSTEMATIC UNCERTAINTIES}
\label{syserr}

The two dominant systematic uncertainties come from the acceptance and 
efficiency correction procedures. 
As described in Section~\ref{effcorr}, acceptance and efficiency
corrections have been computed using MC simulation, with an additional 
correction applied to compensate for MC deficiencies in the low \pt\ 
region between 0.4-0.8 \gevc. The two correction procedures are largely
independent, which allows us to evaluate 
the systematic uncertainties from these two sources separately. 
The details are described below.

\vspace{1em}
1. We study the sensitivity of this measurement to the  
differences between the MC predictions and the shapes of the observed 
\vv\ kinematical distributions.
We use the following two sets of MC events. 
The first is created using the default {\sc pythia} MC.
The second 
is the one used for efficiency studies using the embedding procedure:
the \vvs\ in this set have non-physical distributions roughly uniform
in \pt\ but not in $\eta$. 
The different correction factors evaluated from the above data sets 
are applied to the measured distributions.
Half the difference between the corrected distributions is taken as the 
systematic uncertainty on the distributions themselves,
which amounts to about 10\% for the \kz\ and \lz\ \pt\ distributions, roughly 
constant over the whole spectrum.

The effect on the mean \pt\ value is 3\% for \kz\ and 4\% for \lz.
For the \kz\ and \lz\ multiplicity distributions, the 
systematic variation ranges from 10\% to $\sim$25\%.
As a function of \Nch, the systematic uncertainty 
on the number of \kz\ ranges from a few percent to roughly 20\% at 
the highest charged multiplicities.

\vspace{1em}
2. The systematic uncertainty due to the correction for the undetected \vv\ 
in the \pt\ region between 0.4 and 0.8 \gevc\ has been evaluated in the
following way.
The procedure defined in Section~\ref{effcorr}, point 7, 
is repeated using \pt\ and proper lifetime distributions both corrected
with {\sc pythia} MC and with the embedding-based correction.
The total number of \vv\ is computed by integrating the corrected
\pt\ and lifetime spectra for each of the two cases. We end up
with four different evaluations of the number of undetected \vv.
By comparing the numbers obtained from all combinations, we observe that 
the largest
difference amounts to about 50\% of the correction value. This number is
taken as the systematic uncertainty on this correction and is counted as
a contribution to the systematics on the the total number of measured \kl.

\begin{figure}{
\scalebox{1}{
\includegraphics[1,2][423,254]{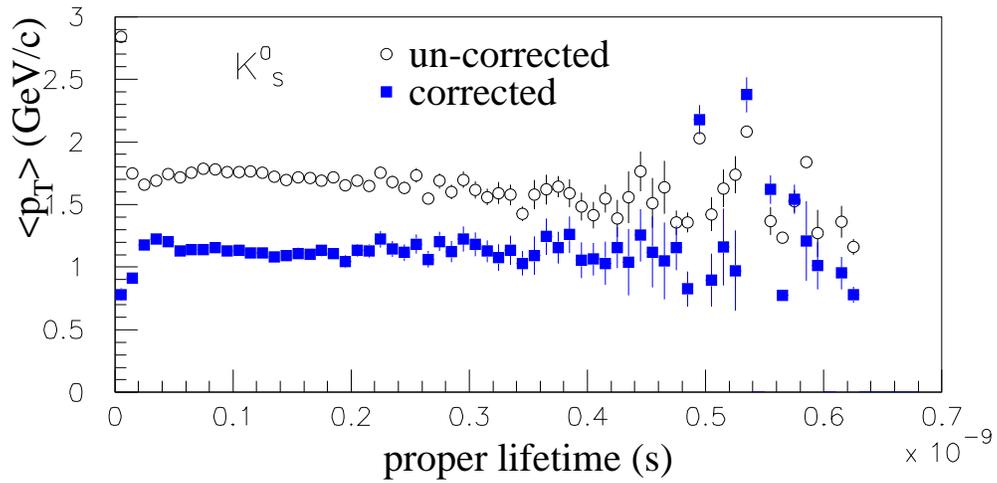}}
\vspace{-1ex}
\scalebox{1}{
\includegraphics[1,2][423,254]{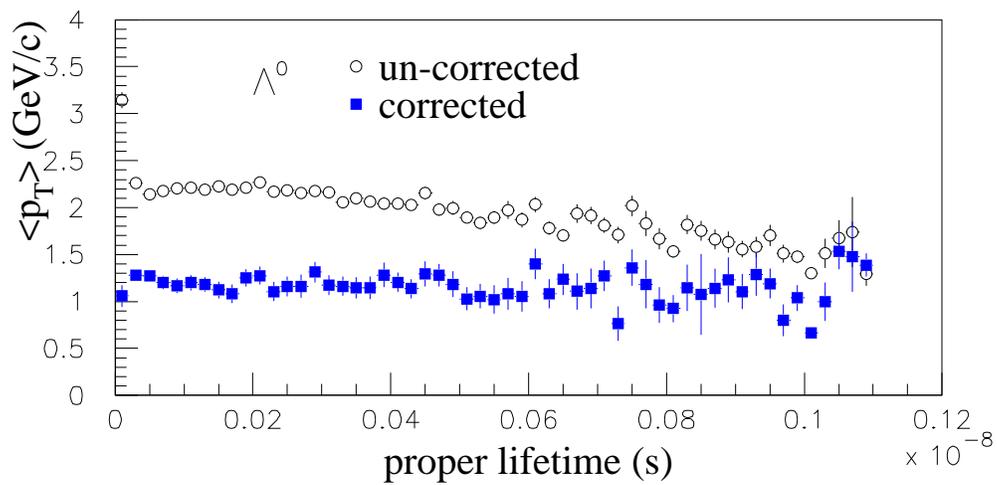}} }
\caption{\label{fig:ptmlife}The mean \pt\ of \kz\ and \lz\ as a function of 
  proper lifetime ($t$) at 1800 GeV. Raw and corrected data are shown.}
\end{figure}

\begin{figure}
\scalebox{1.0}{
\includegraphics[1,20][306,298]{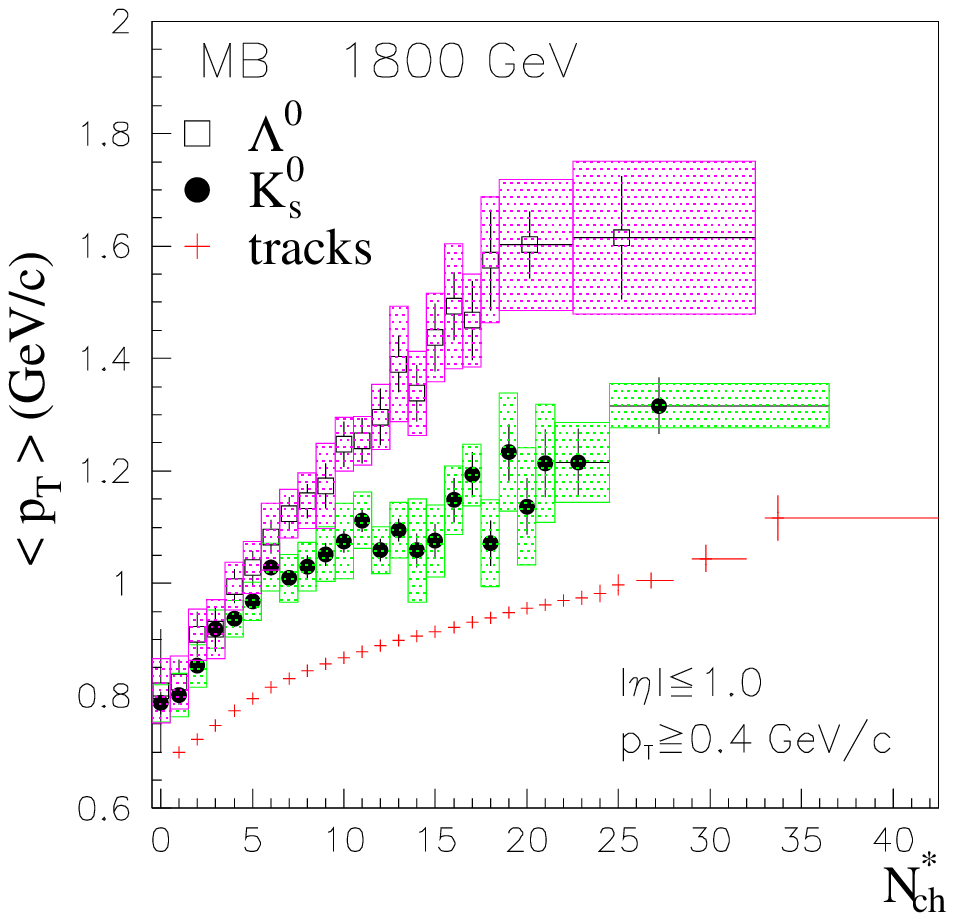}}
\caption{\label{fig:meanpt_mb_1800}Average transverse momentum \ptm\ of \kz\
  and \lz\ at 1800 GeV
  as a function of the event charged multiplicity (\Nch).  MB data are shown.
  For comparison, the mean \pt\ of charged particles measured in the same
  phase space region is also plotted~\cite{noi}. The filled squares around the
  points delimit the systematic uncertainties.}
\end{figure}

\begin{figure}
\scalebox{1.0}{
\includegraphics[1,20][306,298]{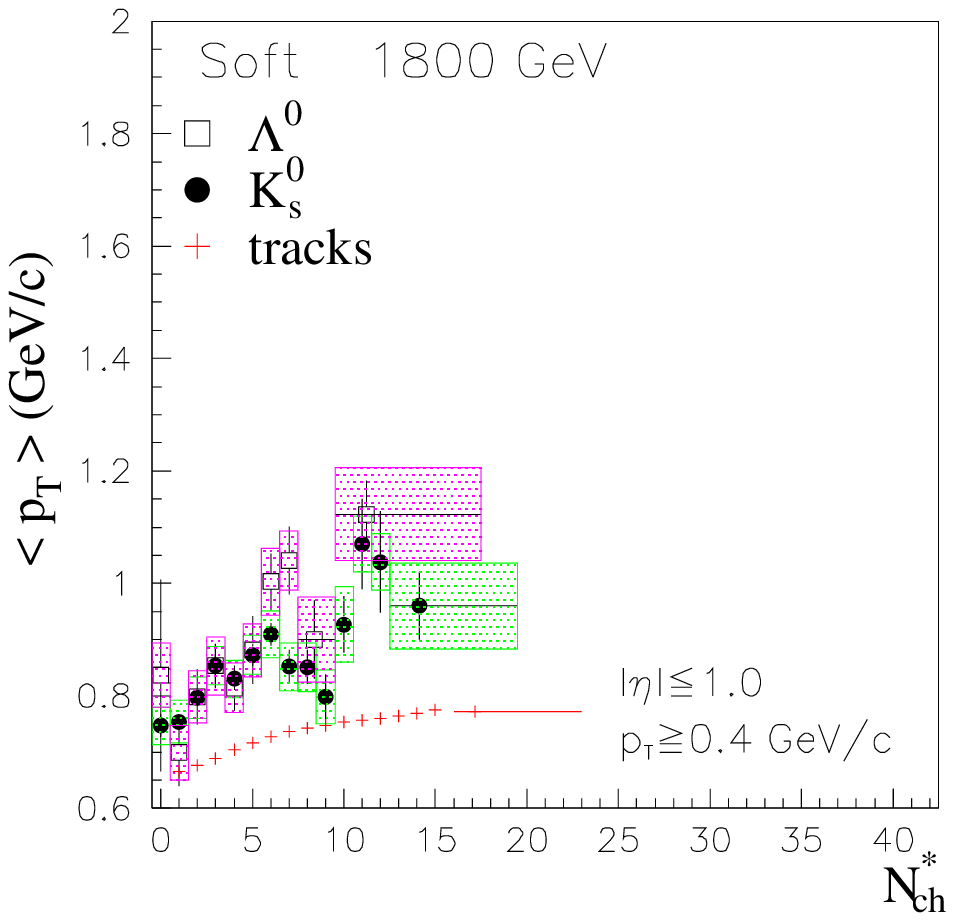}}
\caption{\label{fig:meanpt_nj_1800}Average transverse momentum \ptm\ of \kz\
  and \lz\ at 1800 GeV
  as a function of the event charged multiplicity (\Nch). $Soft$ data are shown.
  For comparison, the mean \pt\ of charged particles measured in the same
  phase space region is also plotted~\cite{noi}. The filled squares around the
  points delimit the systematic uncertainties.}
\end{figure}

\begin{figure}
\scalebox{1.0}{
\includegraphics[1,20][306,298]{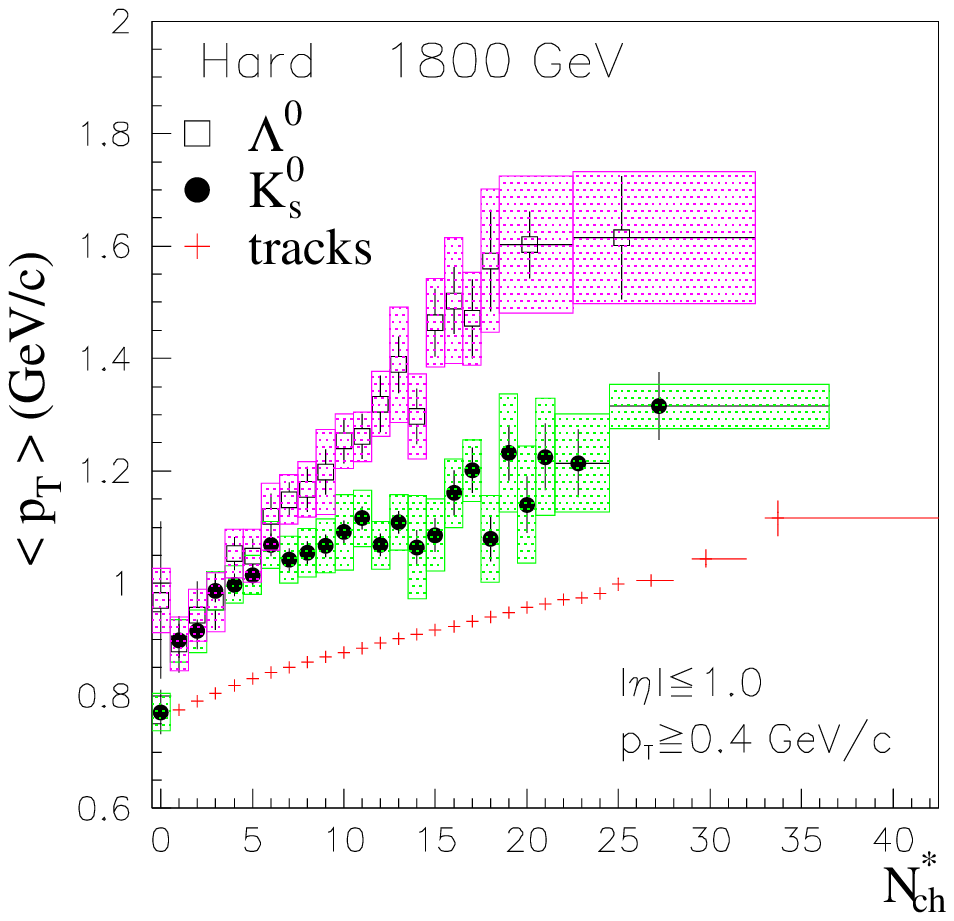}}
\caption{\label{fig:meanpt_j_1800}Average transverse momentum \ptm\ of \kz\
  and \lz\ at 1800 GeV
  as a function of the event charged multiplicity (\Nch). $Hard$ data are shown.
  For comparison, the mean \pt\ of charged particles measured in the same
  phase space region is also plotted~\cite{noi}. The filled squares around the
  points delimit the systematic uncertainties.}
\end{figure}

\begin{figure}
\scalebox{1.0}{
\includegraphics[1,20][306,298]{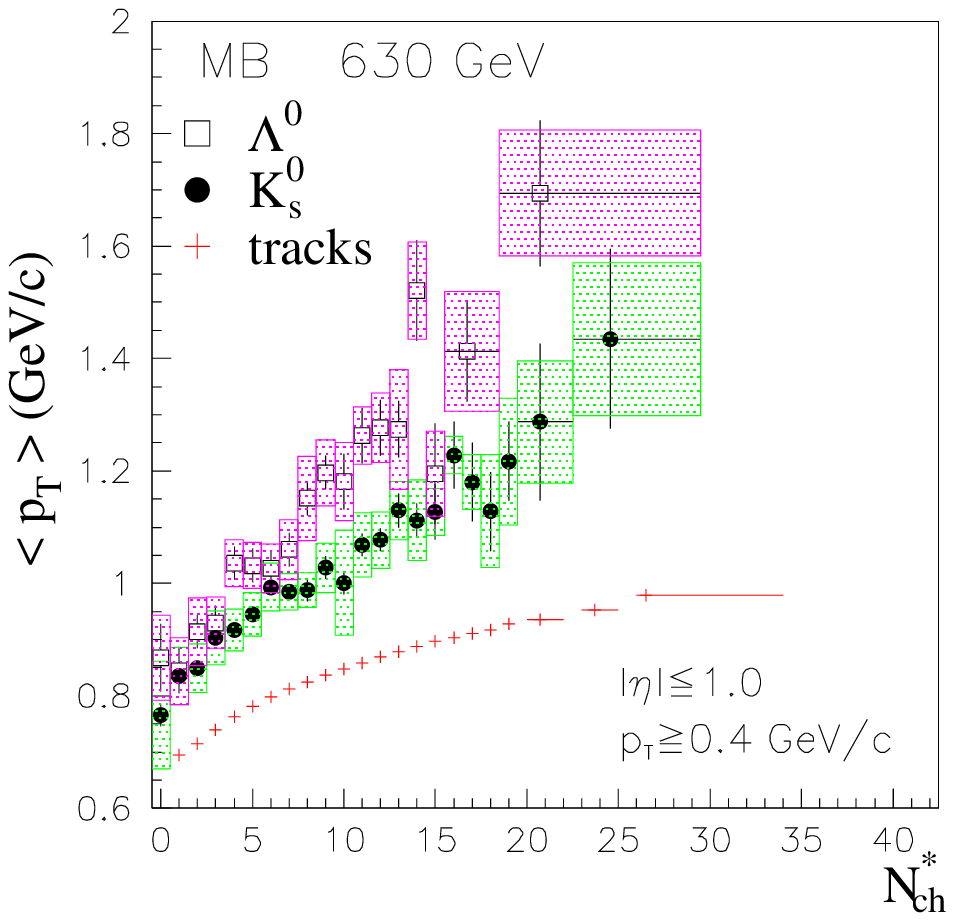}}
\caption{\label{fig:meanpt_mb_630}Average transverse momentum \ptm\ of \kz\
  and \lz\ at 630 GeV
  as a function of the event charged multiplicity (\Nch).  MB data are shown.
  For comparison, the mean \pt\ of charged particles measured in the same
  phase space region is also plotted~\cite{noi}. The filled squares around the
  points delimit the systematic uncertainties.}
\end{figure}

\begin{figure}
\scalebox{1.0}{
\includegraphics[1,20][306,298]{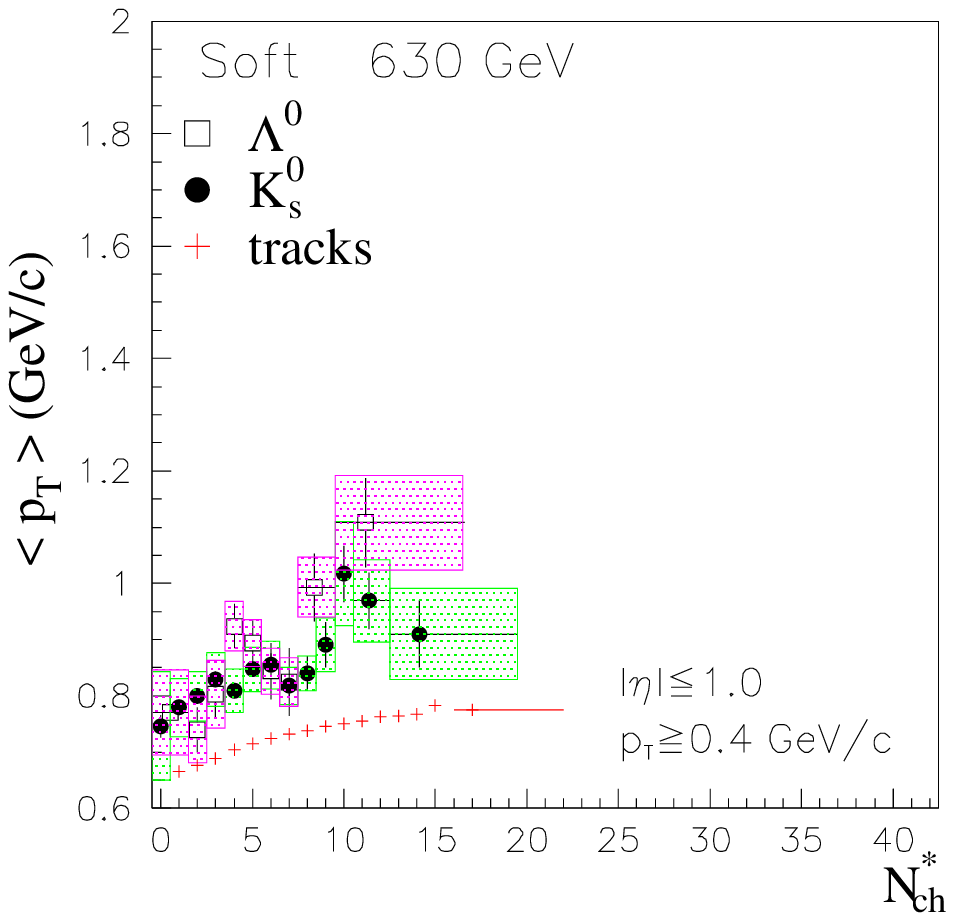}}
\caption{\label{fig:meanpt_nj_630}Average transverse momentum \ptm\ of \kz\
  and \lz\ at 630 GeV
  as a function of the event charged multiplicity (\Nch). $Soft$ data are shown.
  For comparison, the mean \pt\ of charged particles measured in the same
  phase space region is also plotted~\cite{noi}. The filled squares around the
  points delimit the systematic uncertainties.}
\end{figure}

\begin{figure}
\scalebox{1.0}{
\includegraphics[1,20][306,298]{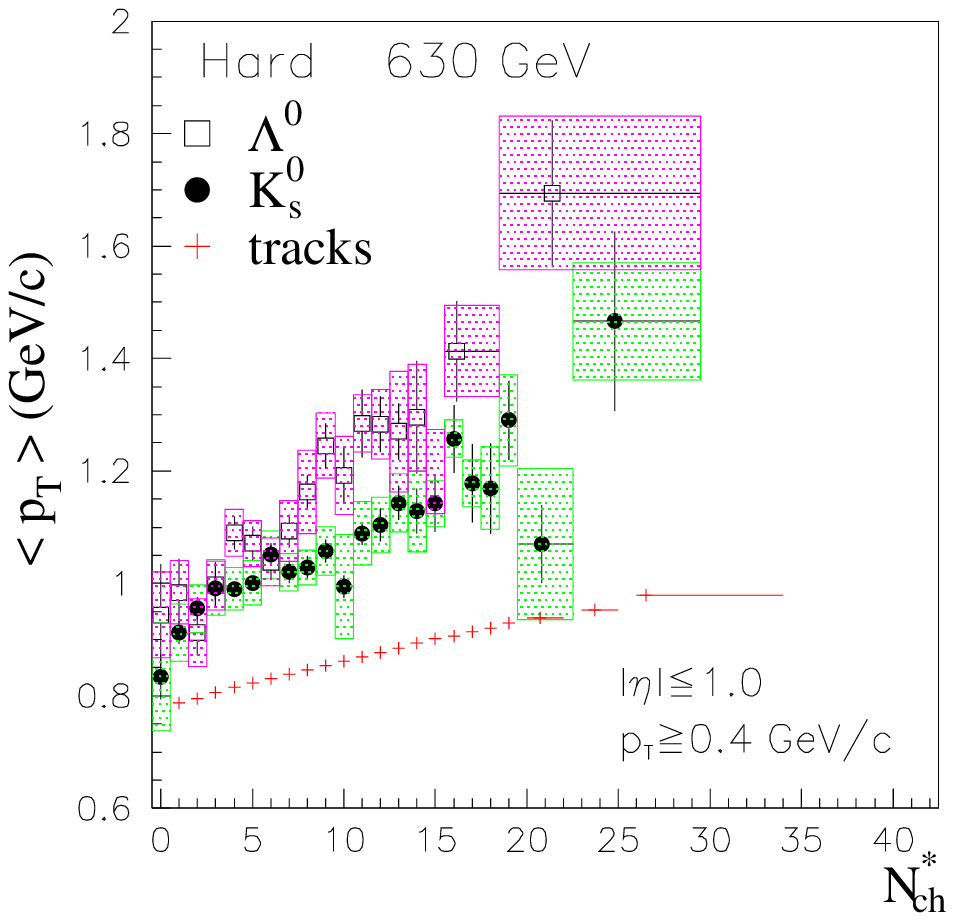}}
\caption{\label{fig:meanpt_j_630}Average transverse momentum \ptm\ of \kz\
  and \lz\ at 630 GeV
  as a function of the event charged multiplicity (\Nch). $Hard$ data are shown.
  For comparison, the mean \pt\ of charged particles measured in the same
  phase space region is also plotted~\cite{noi}. The filled squares around the
  points delimit the systematic uncertainties.}
\end{figure}

The mean \pt\ values at fixed multiplicities are also affected by the 
correction for the undetected \kl\ in the low \pt\ region.
The systematic uncertainty on the correction is estimated as follows.
First it has been verified that the mean \pt\
after correction is independent of the \kl\ proper lifetime 
in the region used in this analysis (see Fig.~\ref{fig:ptmlife}).
Then, starting with \pt\ distributions at fixed charged multiplicity 
for the subset of 
events with \kl\ proper lifetime greater than $0.8\times10^{-10} s$ 
($1.0\times10^{-10} s$),
the mean \pt\ is computed the same way as described
in Section~\ref{effcorr} and
the difference between the mean \pt\ values for the full dataset and
the high $\tau$ subset is assigned 
as a systematic uncertainty for this correction. 
Since the correction is applied only to calculations of the mean \pt\ at 
fixed multiplicity and of the number of \kl, the systematic 
uncertainty associated with it affects only these measurements. 
It amounts to about 6\% (10\%) for the total number of \kz\ (\lz) and 
affects the average number of \kl\ as a function 
of the charged multiplicity by the same amount.  
These systematic uncertainties combined in quadrature with the other 
systematic uncertainties discussed in this section are included in 
Figures~\ref{fig:meanpt_mb_1800} to \ref{fig:meanpt_j_630}.
\vspace{1em}

3. To investigate the systematic effect of the track reconstruction procedure
on the efficiency correction, we compare our result with a set of MC events 
where the tracks 
are reconstructed using the CTC information alone, as opposed to the
default SVX+CTC track reconstruction.
We find that the variation on the final corrected \pt\ distribution is
negligible.
\vspace{1em}

4. Other sources of systematic uncertainties include the dependence of the 
results on the instantaneous luminosity and the uncertainty associated
with the identification and selection of good isolated
\pbarp\  interactions from secondary or closely spaced event 
vertices (see~\cite{noi}). 
The first may affect the results because higher
luminosity gives higher detector occupancy which in turn can alter the \vv\
identification. This has been investigated by analyzing
data samples recorded at different instantaneous luminosities. 
The results show no observable effect. 
The second source can lead to incorrect event selection and produce 
associations of tracks that fake a \vv.
This source has been investigated by
comparing data samples with different requirements for a good \pbarp\  
vertex~\cite{noi}. 
The results give systematic variations smaller than 9\%
on the overall number of \kl.
\vspace{1em}

\begin{figure}
\scalebox{1.0}{
\includegraphics[1,20][306,298]{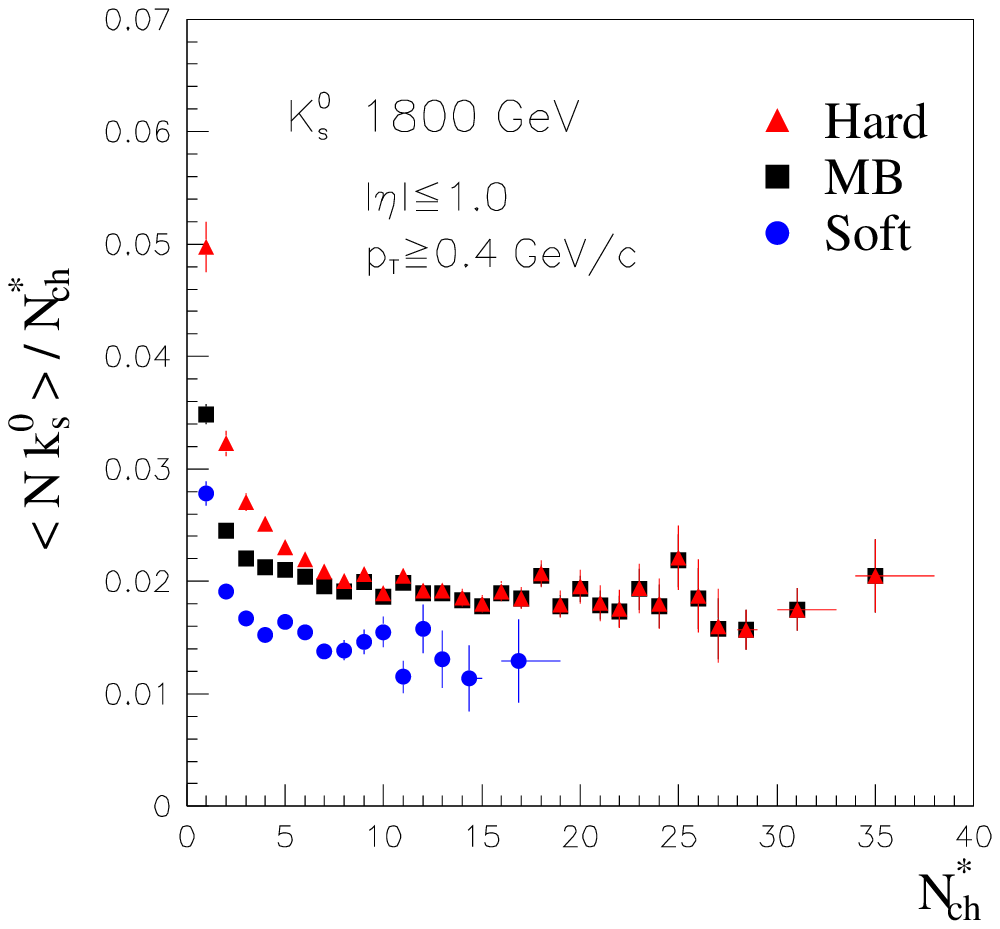}}
\caption{\label{fig:nvsm_k0_1800}Mean number of \kz\ per event divided by the
      charged multiplicity (\Nch) and plotted as a function of \Nch.
      The MB, $soft$ and $hard$ data at 1800 GeV are shown.}
\end{figure}

\begin{figure}
\scalebox{1.0}{
\includegraphics[1,20][306,298]{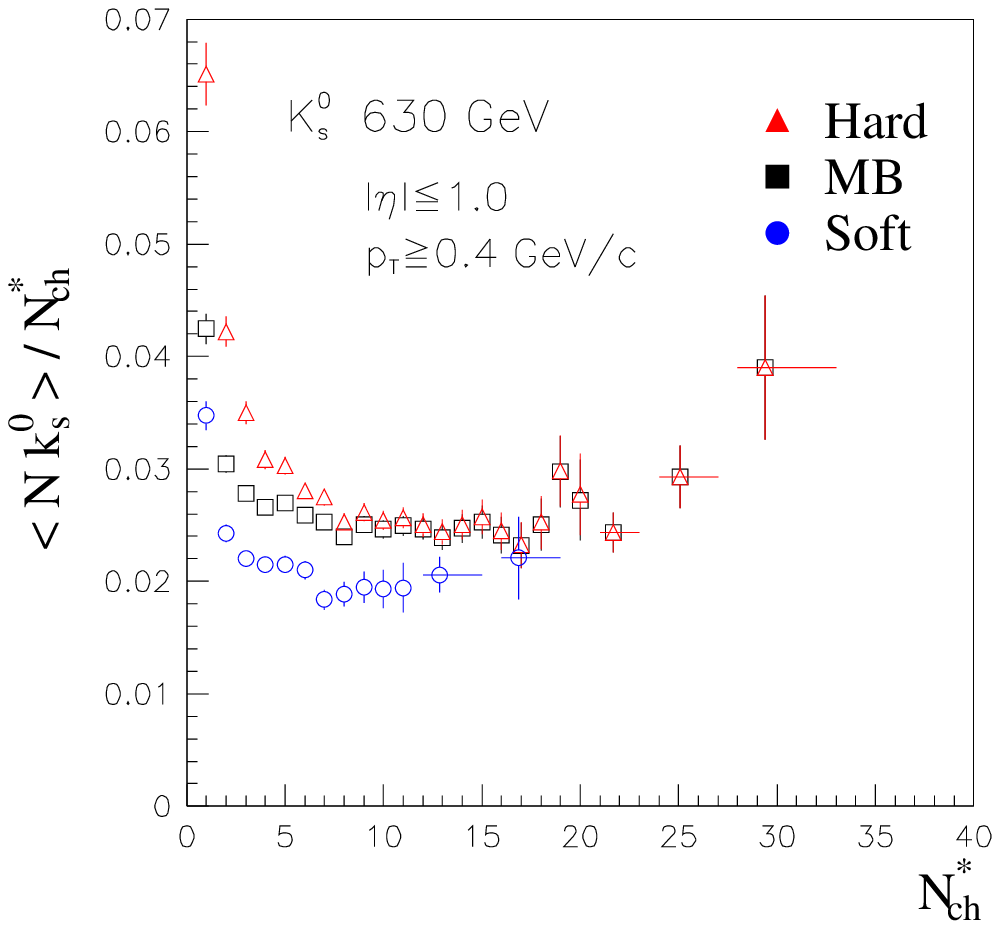}}
\caption{\label{fig:nvsm_k0_630} Mean number of \kz\ per event divided
 by the charged \m\ (\Nch) and plotted as a function of \Nch.
 The MB, $soft$ and $hard$ data at 1800 GeV are shown.}
\end{figure}

\begin{figure}
\scalebox{1.0}{
\includegraphics[1,20][306,298]{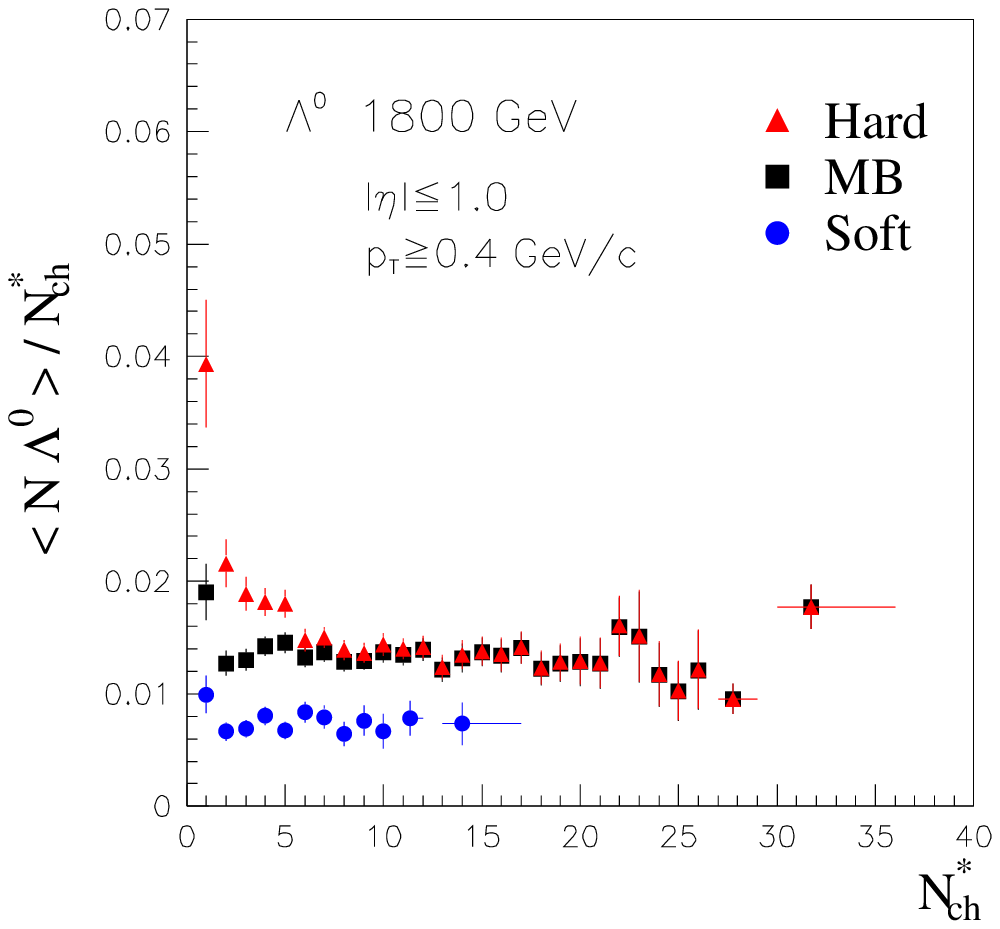}}
\caption{\label{fig:nvsm_l0_1800}Mean number of \lz\ per event divided by the
      charged multiplicity (\Nch) and plotted as a function of \Nch.
      The MB, $soft$ and $hard$ data at 1800 GeV are shown.}
\end{figure}

\begin{figure}
\scalebox{1.0}{
\includegraphics[1,20][306,298]{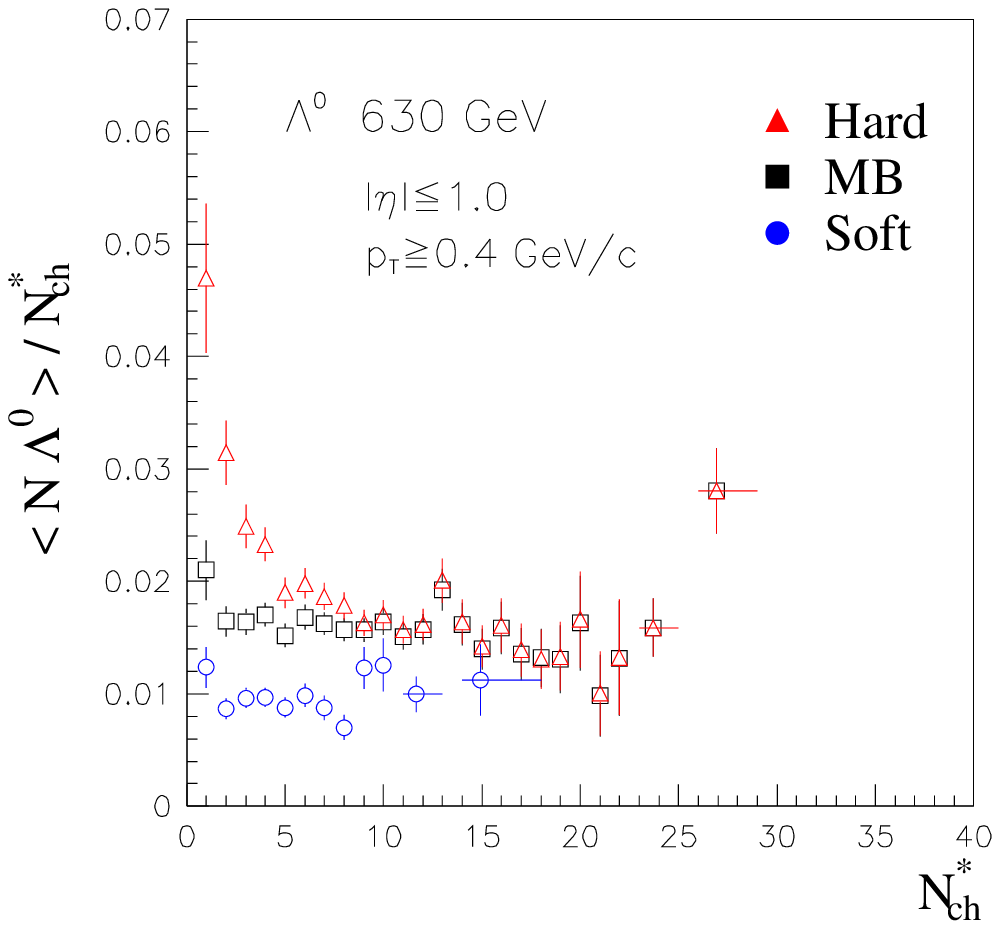}}
\caption{\label{fig:nvsm_l0_630}Mean number of \lz\ per event divided by the 
      charged multiplicity (\Nch) and plotted as a function of \Nch.
      The MB, $soft$ and $hard$ data at 630 GeV are shown.}
\end{figure}

5. The uncertainty on the correction is distributed in different ways for 
different observables. As a consequence, the integral
of the corrected distribution of each variable is different.
For example, the total number of \kl\ extracted from the 
integral of the corrected \pt\ distribution may be very different from
that extracted from the multiplicity distribution.
In particular, as discussed in the previous section, the \pt\ 
correction has been observed to be unreliable for \pt\ \les\ 0.7 \gevc\
where a large part of the \vv\ cross-section lies, so that the
area under the distribution may be subject to large uncertainties.
Given this, we use the global (integrated) correction from the {\sc pythia} MC 
as a correction factor for the total number of \kl.
We renormalize each distribution to this number to which we attribute a 30\%
systematic uncertainty. This value is determined as the maximum difference
that was found between the global corrected number of \vv\ and the integral 
of any corrected distribution.
Such uncertainty reflects on the \kz/$\pi$ ratios
and on the absolute scale of the ratios of the mean number
of \kl\ to the charged multiplicity plotted in Figures~\ref{fig:nvsm_k0_1800}
to \ref{fig:nvsm_l0_630}.
\par\noindent
Table~\ref{tab3} reports a summary of all the systematic uncertainties 
discussed.

\section{ANALYSIS RESULTS}
\label{resdis}
\subsection{Results}
\label{incldis}

All data presented are subject to \pt\ \gt\ 0.4 \gevc\ and 
$|\eta| \leq$ 1 requirements, as specified in Section~\ref{k0sel}, and
are corrected for acceptance and vertex-finding efficiency.
Systematic uncertainties are not included except where explicitly stated.
Table~\ref{tab1} shows the raw and corrected numbers of  \kl\ selected 
in our fiducial region for the full MB sample as well as for 
the {\it soft} and 
the {\it hard} samples. The corrected mean number of \kl\ per event in each 
sample is also shown; systematic uncertainties are included. 

\begin{figure}
\scalebox{1}{
\includegraphics[130,205][430,610]{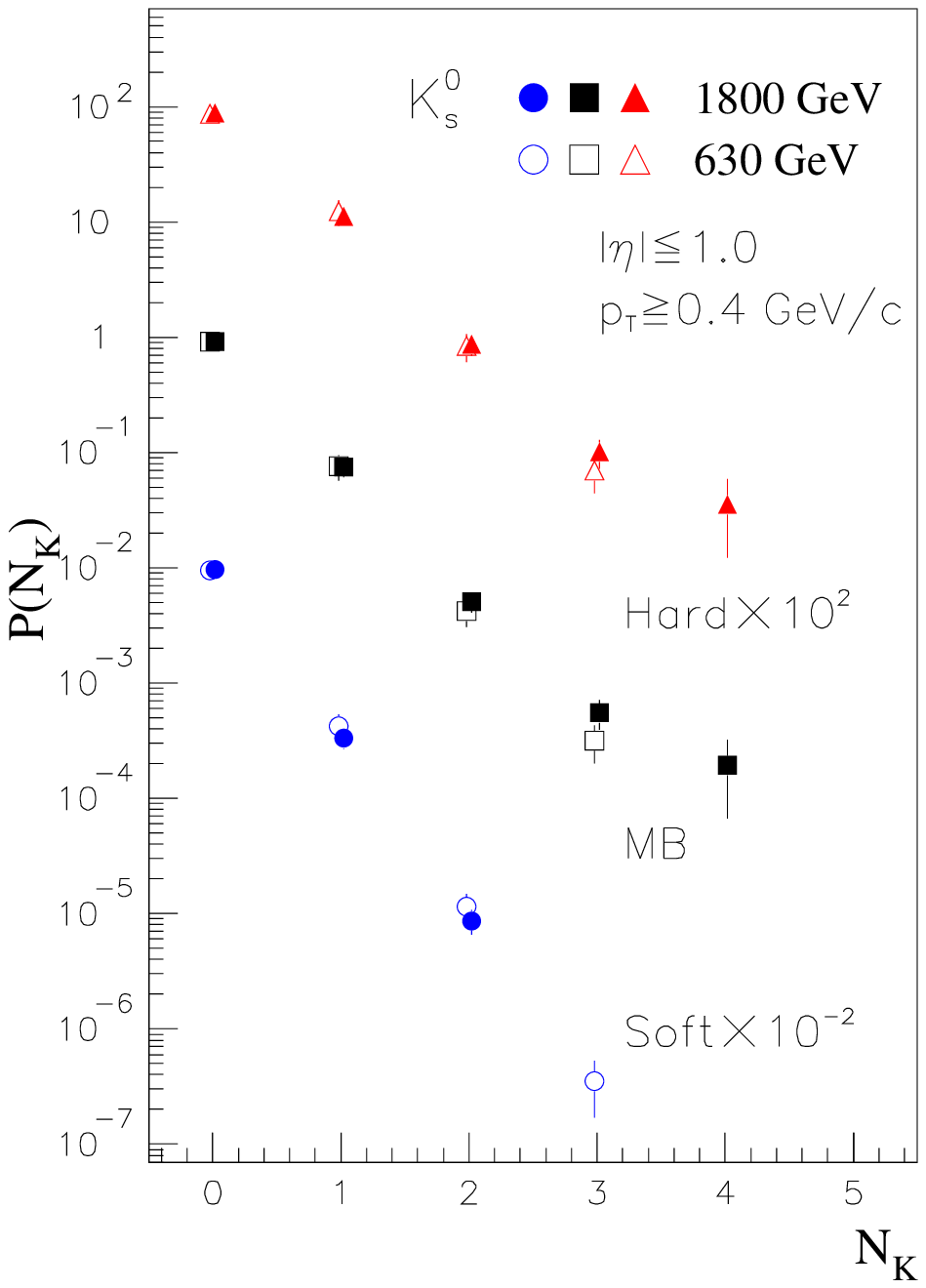}}
\caption{\label{fig:mult_k0}Distribution of the multiplicity of \kz\ at 
  1800 (full symbols) and 630 GeV (open symbols). 
  P($N_{K}$) = (number of events with $N_{K}$ \kz)/(total number of events).
  MB, $soft$ (divided by 100) and $hard$ data (multiplied by 100) are shown.}
\end{figure}

\begin{figure}
\scalebox{1}{ 
\includegraphics[130,205][430,610]{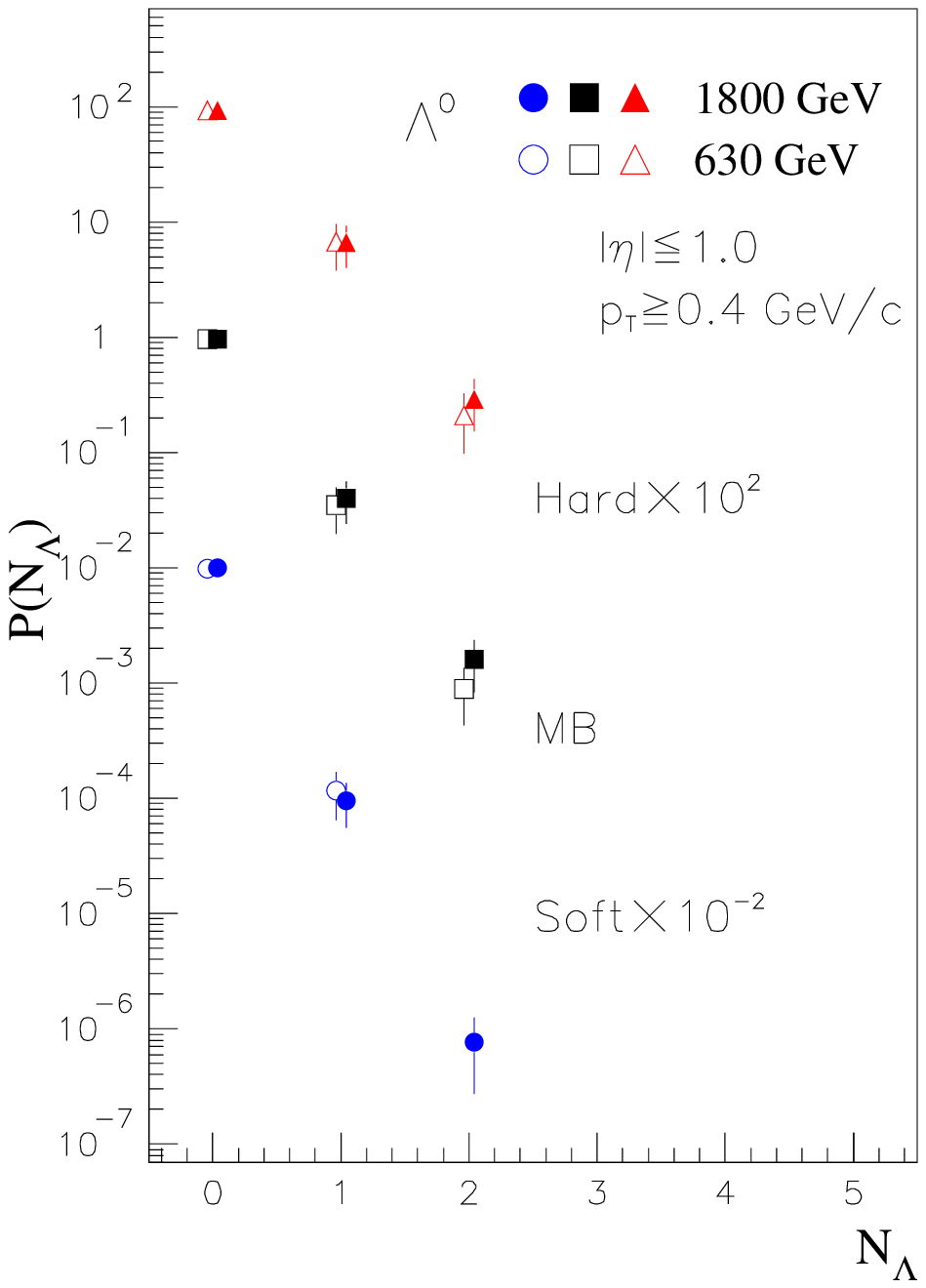}}
\caption{\label{fig:mult_l0}Distribution of the multiplicity of \lz\ at 
   1800 (full symbols) and 630 GeV (open symbols).
P($N_{\Lambda}$) = (number of events with $N_{\Lambda}$ \lz)/(total number 
of events). 
   MB, $soft$ (divided by 100) and $hard$ data (multiplied by 100) are shown.}
\end{figure}

In Fig.~\ref{fig:mult_k0} for the \kz\ and in Fig.~\ref{fig:mult_l0} for
the \lz, the 
normalized multiplicity of \kl\ for the MB, {\it soft} and {\it hard} 
events is shown separately for the $\sqrt{s}$=1800 GeV (solid symbols) 
and 630 GeV (open symbols) data.
The probability of producing one or more \lz\ is lower than the equivalent 
\kz\ probability, and the difference increases with \vv\
multiplicity.
This behavior is more pronounced in the $soft$ subsample.
The results shown in Figures~\ref{fig:mult_k0} and \ref{fig:mult_l0}, 
with their statistical errors, are reported in Table~\ref{tab4}.

\begin{figure}
\scalebox{1.0}{
\includegraphics[130,205][430,610]{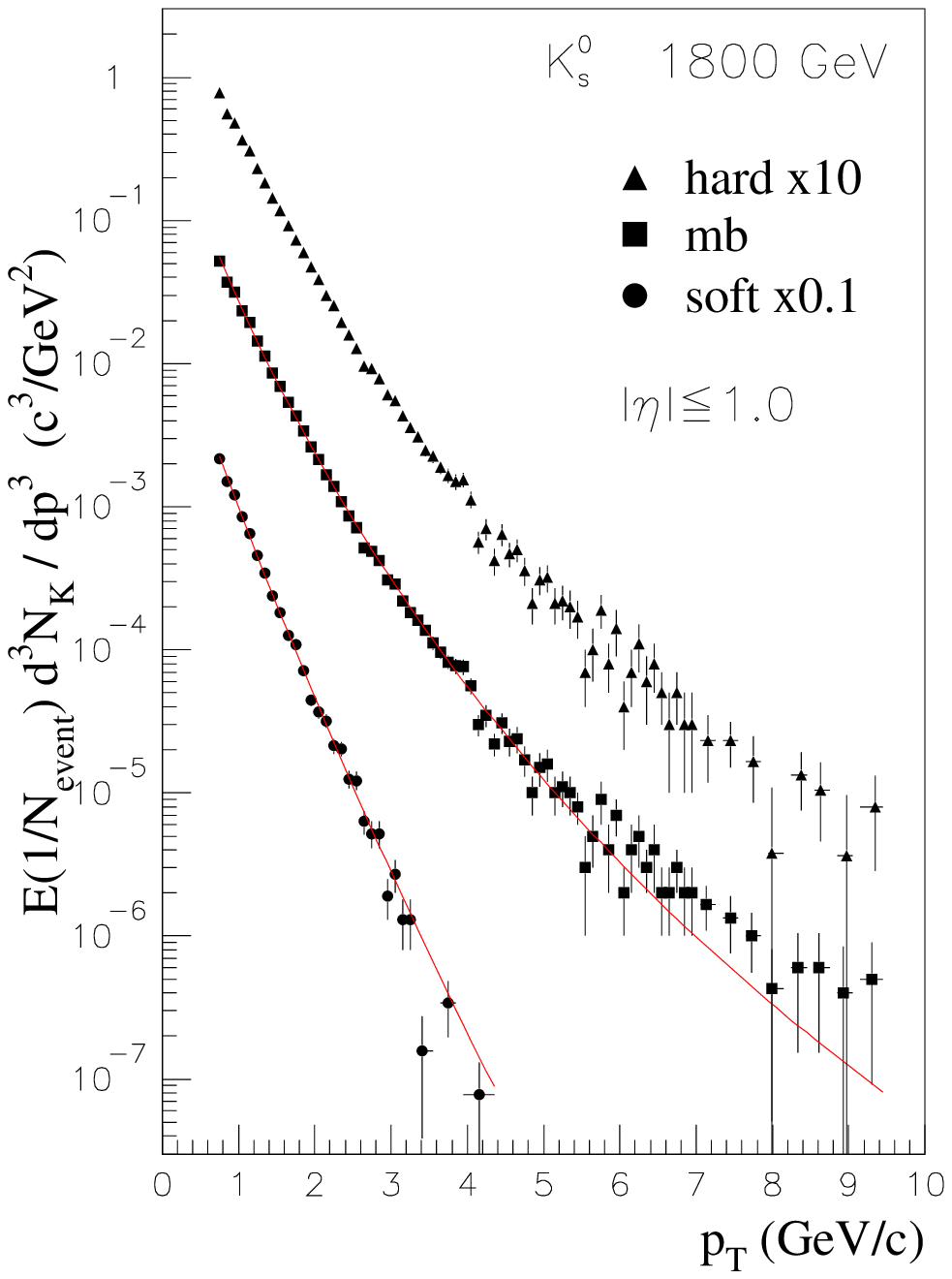}}
\caption{\label{fig:pti_k0_1800}\kz\ inclusive invariant \pt\ distributions at 
    1800 GeV. MB, $soft$ and $hard$ data are shown, 
    normalized to the number of events in each sample.
    E is the particle energy and $N_{event}$ is the total number of events
    which contribute to the distribution. 
    To separate the curves, $hard$ data points are multiplied by 10 
    and $soft$ data points by 0.1.
    The solid lines represent the best fits to Eq.(\ref{eq:fpt2}) of the text.}
\end{figure}

\begin{figure}
\scalebox{1.0}{
\includegraphics[130,205][430,610]{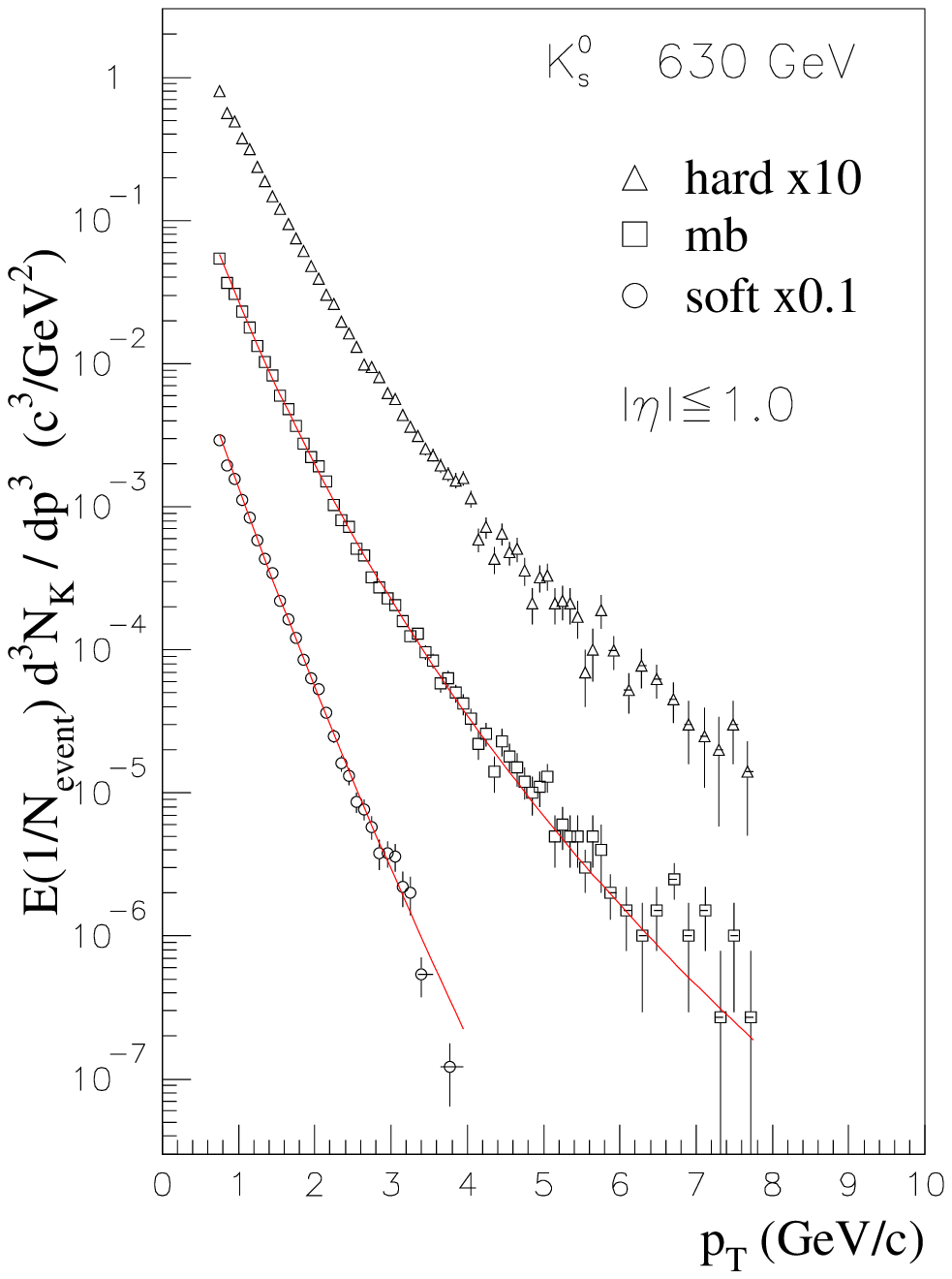}}
\caption{\label{fig:pti_k0_630}\kz\ inclusive invariant \pt\ distributions at 
    630 GeV MB, $soft$ and $hard$ data are shown, 
    normalized to the number of events in each sample. 
    E is the particle energy and $N_{event}$ is the total number of events
    which contribute to the distribution. 
    To separate the curves, $hard$ data points are multiplied by 10 
    and $soft$ data points by 0.1.
    The solid lines represent the best fits to Eq.(\ref{eq:fpt2}) of the text.}
\end{figure}

\begin{figure}
\scalebox{1.0}{
\includegraphics[130,205][430,610]{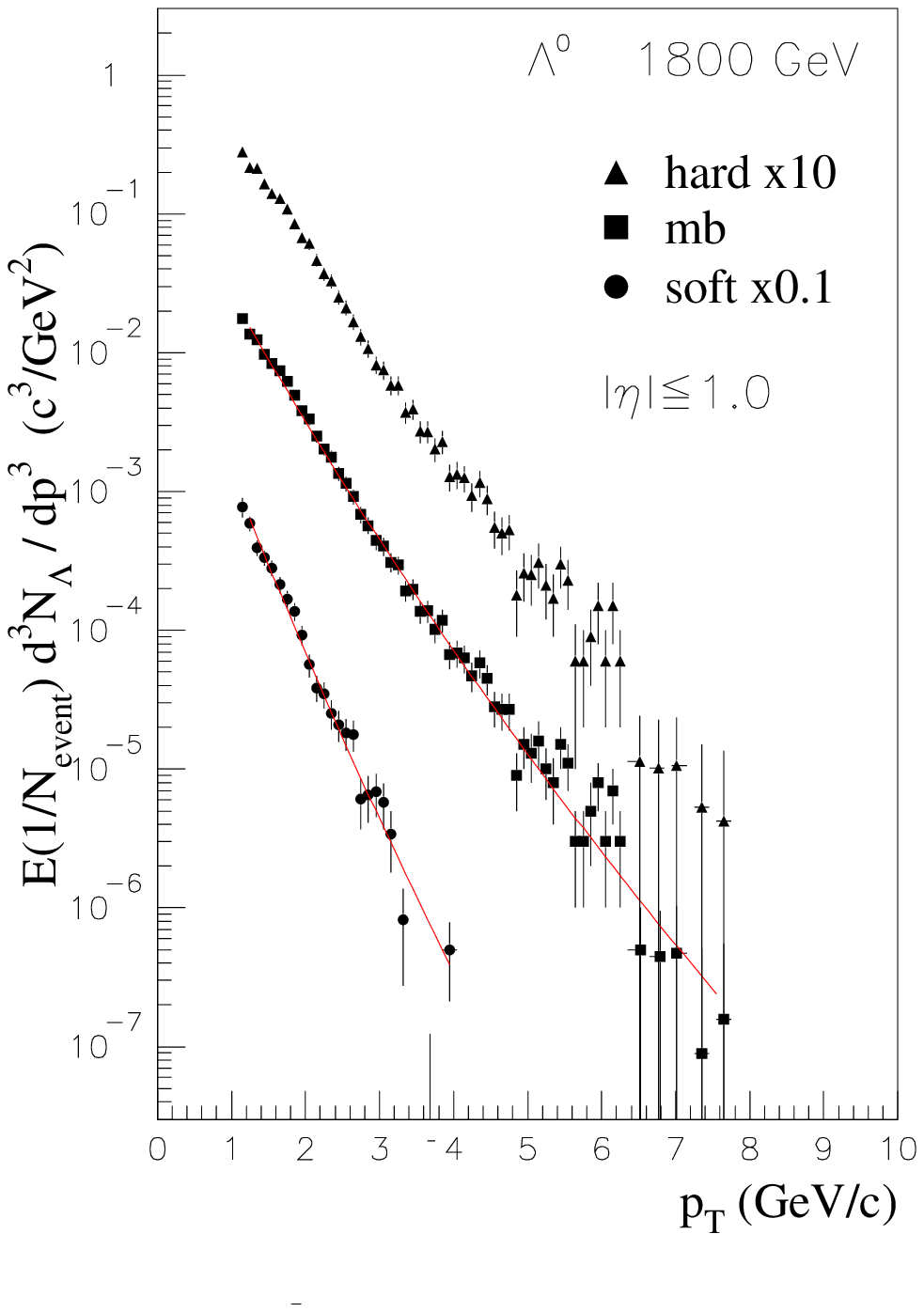}}
\caption{\label{fig:pti_l0_1800}\lz\ inclusive invariant \pt\ distributions at
    1800 GeV. MB, $soft$ and $hard$ data are shown, 
    normalized to the number of events in each sample. 
    E is the particle energy and $N_{event}$ is the total number of events
    which contribute to the distribution.     
    To separate the curves, $hard$ data points are multiplied by 10
    and $soft$ data points by 0.1.
    The solid lines represent the best fits to Eq.(\ref{eq:fpt2}) of the text.}
\end{figure}

\begin{figure}
\scalebox{1.0}{
\includegraphics[130,205][430,610]{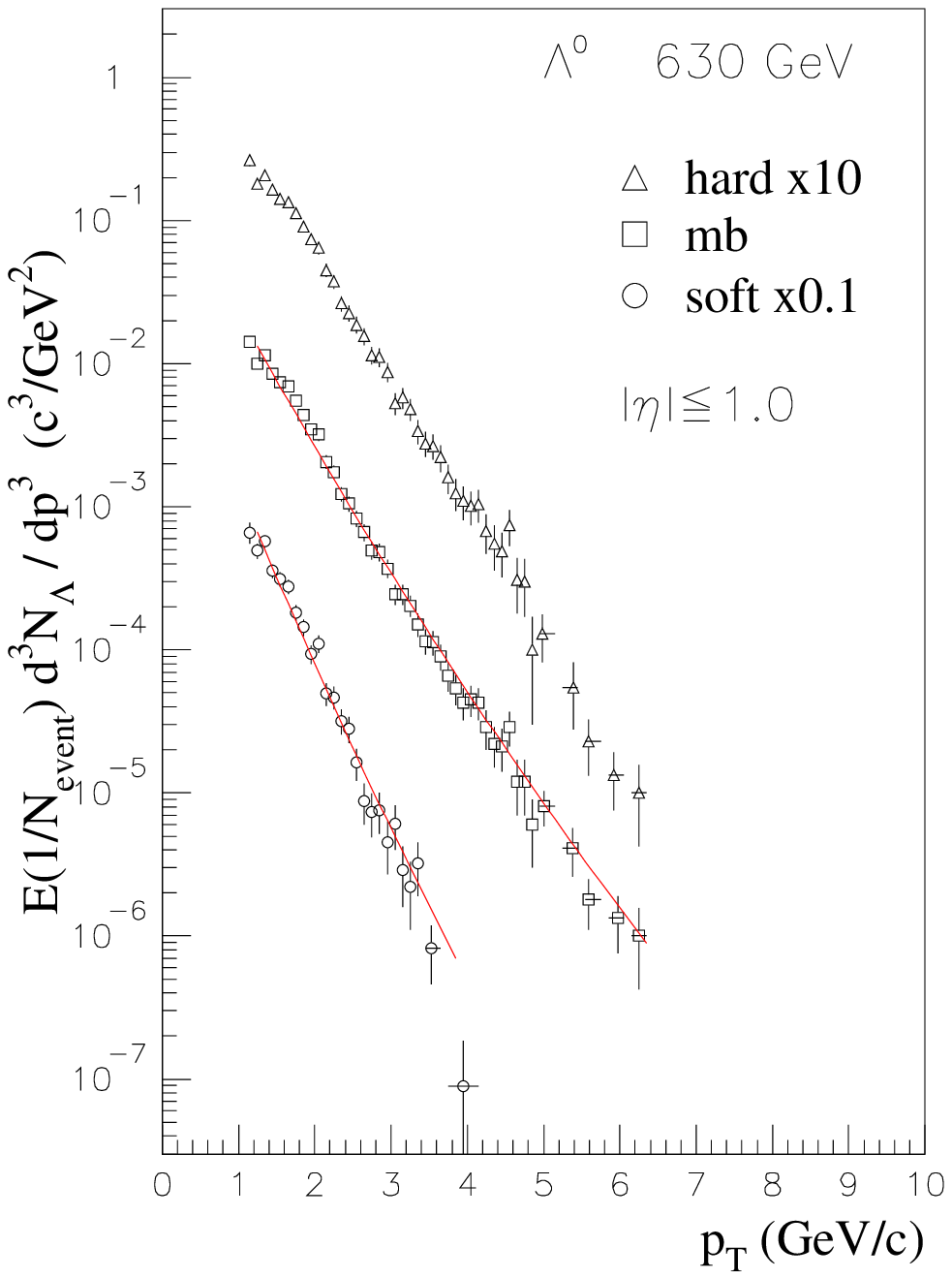}}
\caption{\label{fig:pti_l0_630}\lz\ inclusive invariant \pt\ distributions at
    630 GeV. MB, $soft$ and $hard$ data are shown, 
    normalized to the number of events in each sample.
    E is the particle energy and $N_{event}$ is the total number of events
    which contribute to the distribution. 
    To separate the curves, $hard$ data points are multiplied by 10
    and $soft$ data points by 0.1.
    The solid lines represent the best fits to Eq.(\ref{eq:fpt2}) of the text.}
\end{figure}

The invariant \pt\ inclusive distributions of 
\kz\ are shown in Figures ~\ref{fig:pti_k0_1800} and \ref{fig:pti_k0_630}
at the two energies for the full MB, $soft$ and $hard$ samples.
Data are normalized to the number of events in each sample.
Figures~\ref{fig:pti_l0_1800} and \ref{fig:pti_l0_630} show
the same \pt\ distributions for the \lz.

The dependence of the \kz\ and \lz\ average \pt, calculated as described
in Eq.~(\ref{eq:fpt3}), on the event charged 
multiplicity is shown in 
Figures~\ref{fig:meanpt_mb_1800}~-~\ref{fig:meanpt_j_1800} (1800 GeV)
and ~\ref{fig:meanpt_mb_630}~-~\ref{fig:meanpt_j_630} (630 GeV).
The mean \pt\ of primary charged tracks measured in the same
phase space region, as published in~\cite{noi}, is also shown for 
comparison.
For the \kz\ dataset,
in the region ranging from 0.4 to 0.8~\gevc, the corrected data 
points are assumed to lay on a curve of form (\ref{eq:fpt2}) 
extrapolated from the fit to the measured data points in the region
$p_{T}>$0.8~\gevc\  (details of the correction procedure are described 
in Section~\ref{effcorr}).
Note, that with the kinematical selection used in this analysis, no 
events with \pt(\lz)$<$1~\gevc\ were observed.
For the measurement of \ptm, we fit the spectrum in the region
of \pt$>$1.1~\gevc\ using Eq.(\ref{eq:fpt2}) and extrapolate down to
\pt=0.4~\gevc. We define the \ptm\ as the mean value of the fitted
function.
This definition is adopted in order to compare the \ptm\
with that of \kz\ and of charged tracks.

Figures~\ref{fig:nvsm_k0_1800} to~\ref{fig:nvsm_l0_630}
show the ratio of the mean number of \kl\ per event to the multiplicity 
as a function of the multiplicity itself.
The charged particle multiplicity \Nch\ was chosen as the reference variable 
to analyze \vv\
production. The reason for this choice is based on the observation that the
event charged multiplicity is a global event variable characterizing the
whole multiparticle final state and is related  to the hardness of the 
interaction (see~\cite{noi},~\cite{vanH},~\cite{hwa}).
As in the case of charged particles, possible new structures in the \vv\ 
final state correlations would be exhibited as a function of \Nch.
The dependence of the average \pt\ on multiplicity, for example, remains 
unexplained in any of the current models.

\subsection{Dependence on \Et\ threshold}
\label{depet}

It has been remarked in the previous sections that the identification
of $soft$ and $hard$ events is essentially a matter of definition. In order
to investigate the sensitivity of the above results to the the cluster energy
threshold used to separate $soft$ and $hard$ events, the analysis has been
repeated changing the \Et\ threshold from 1.1 to 3.0 GeV. Although, as 
expected, the higher threshold value influences the global statistics of 
the $soft$ and $hard$ components, 
it preserves the shapes of the inclusive \pt\ distributions and the 
characteristics of the $hard$ and the $soft$ 
samples, and it does not change the shape of the correlations. 
With the new \Et\ threshold the fraction of \kz\ per event rises by the 
same amount, around 30\%, in 
the two samples. This means that the ratio of the rate of \kz\ in $soft$ events
to the same rate in $hard$ events is not influenced by the higher threshold. 

\subsection{Analysis Discussion}
\label{andis}

Some simple observations can be made about Table~\ref{tab1}.  
The fraction of the total \kz\ that falls into the $soft$ subsample is
rather small, ranging from about 
30\% at 630 GeV to about 18\% at 1800 GeV (19\% and 10\% for 
\lz\ respectively). 
The corrected mean number of \kz\ produced per event in the full MB sample 
is about (8.6$\pm$2.6)\% at 630 GeV and (8.8$\pm$2.6)\% at 1800 GeV 
(respectively (3.7$\pm$1.1)\% and (4.3$\pm$1.0)\% for \lz).

The \kz/$\pi$ cross-section ratio may be obtained by fitting the \kz\ and
charged-track invariant \pt\ distributions in the available \pt\ range and
extrapolating the fitted functions down to the minimum \pt\ value.
This ratio is evaluated both for \pt$_{min}$=0 \gevc\ and 
\pt$_{min}$=0.4 \gevc. 
With the above technique, a ratio of 0.13$\pm$0.04
(including the systematic uncertainty) at $\sqrt{s}$=1800 GeV and 
0.18$\pm$0.05 at 630 GeV is obtained for \pt$_{min}$=0.4 \gevc.
The same ratio for \pt$_{min}$=0 \gevc, gives 0.14$\pm$0.05
at $\sqrt{s}$=1800 GeV and 0.19$\pm$0.06 at 630 GeV (Table~\ref{tab2}).  
These last measurements are compatible with the previous CDF results
\cite{CDFO}, though are slightly higher at $\sqrt{s}=630$~GeV.
\noindent
In Table~\ref{tab2}, the corresponding values for 
the $soft$ and $hard$ subsamples are reported. It is remarkable that \kz/$\pi$
ratio is about two times larger in $soft$ than in MB events.

Studies of the production of strange particles \kz\ and \lz\ in 
proton-antiproton interactions at different $\sqrt{s}$ are described in
Ref.~\cite{UA51} at $\sqrt{s}$=540 GeV and ~\cite{UA52} at 200 and 
900 GeV.
In Refs.~\cite{CDFO}, \cite{E735}, \cite{alex}, results at $\sqrt{s}$=1800 
GeV are presented. Comparison with our results is
restricted to the full MB samples; furthermore,
it should be noted that here no absolute cross-sections 
are provided. Comparison with Refs.~\cite{UA51}-\cite{E735} also 
requires taking 
into account the different \pt\ and $\eta$ regions selected.

\vspace{1ex}
A direct comparison of the invariant \pt\ distribution of \kz\ can 
be done with Ref.~\cite{CDFO}. 
There the \pt\ distribution of \kz\ is fitted to the 
functional power-law form of Eq.(\ref{eq:fpt2}),
fixing the parameter $p^{0}$ to 1.3~\gevc.
The average $\overline{p_{T}}$ is computed from the parameters of the fit
as: 
\begin{equation}
\overline{p_{T}} = 2 \frac{p^{0}}{n - 3}
\label{eq:fpt4}
\end{equation}
With the new increased statistics and larger \pt\ range,
the fit with the $p^{0}$
parameter fixed, while giving a reasonable description of the \pt\
spectrum in the low-\pt\ region, does not describe the data at higher \pt.
It yields a $\overline{p_{T}}$ compatible with the previous one 
(see Table~\ref{tab5}) but with a large \kdof.
The best fit to our distribution (shown in Figures~\ref{fig:pti_k0_1800} to
\ref{fig:pti_l0_630} for the $soft$ and MB samples) is obtained with 
this form when all three parameters are allowed to vary freely and 
the fit region is restricted to \pt\ $>$1~\gevc.
A summary of the results is reported in Table~\ref{tab5}.
The measurements reported in 
Refs.~\cite{UA51}-\cite{alex} were done at 
different energies and in different phase space regions.

From our best fit of MB sample data at 1800 GeV (630 GeV), the mean
\pt\ of \kz\ is (0.75~$\pm$~0.07)~\gevc\ ((0.70~$\pm$~0.08)~\gevc).
These values are significantly higher than the previous CDF measurements
due to the higher statistics in the high \pt\ tail of the distribution.

Taking into account the different conditions and the method of measurement, 
it is possible to compare to UA5 data (Refs.~\cite{UA51},~\cite{UA52}) as well;
our present measurement is also higher in this case.
For completeness, the fit results of the \pt\ invariant distribution 
for the $soft$ subsample are also reported in Table~\ref{tab5}.
A second fitting function used is of the form:
\begin{equation}
 E \frac{d^{3}N_{\it k}}{dp^{3}} = exp (A + B \pt), \ \ \ \mathrm{where}
\ \ \ \overline{p_{T}} = - {\frac{2}{B}} \cdot
\label{eq:fpt5}
\end{equation}
At both energies, we obtain a good \kdof\ using this function
(see Table~\ref{tab5}).
Therefore, the shape of the $soft$ distribution is also well described
by an exponential function; the mean \pt\ of the fit is generally 
larger than what is obtained using Eq.(\ref{eq:fpt2}).
For \lz, a systematically higher mean \pt\ than other 
experiments at equivalent energy is obtained (compare with 
Refs.~\cite{UA52} and \cite{E735}).
In this case as well, MB data can be equally well fitted by 
form~(\ref{eq:fpt2}) and by an exponential function. 
A summary of these results is in Table~\ref{tab6}.

\vspace{1ex}
The increase of the mean \pt\ (computed as in 
Eq.~(\ref{eq:fpt3})) of the observed \kz\ as a function 
of the event charged multiplicity is always larger than that of 
charged tracks. 
The increase for \lz\ is even larger, leading to the 
conclusion that it depends on the particle mass, as expected.
A similar analysis is also reported in Ref.~\cite{alex}. 
A direct comparison is not possible because of the different \pt\ 
range and $\eta$ acceptance, which reflect in larger 
multiplicities.
However, a rise in mean \pt\ with heavier particle masses is clearly observed.

In the analysis of charged tracks~\cite{noi}, all the correlations examined
in the MB and in the $hard$ samples showed different behaviors with respect 
to \ecms, while a clear invariance was seen in the $soft$ sample.
With the available \kl\ statistics it is not possible to discern 
any difference
in the \ptm\ dependence on multiplicity at the two energies, even in the 
full MB sample. 
Nevertheless, the behavior of the three subsamples is clearly different.
We note that the mean \kl\ \pt\ increases with \Nch\ 
also in the $soft$ subsample, a feature that is not explained by the 
current models(\cite{sjo},\cite{noi},\cite{vanH},\cite{hwa},\cite{barsh}).
This observation also holds for charged hadrons, as discussed 
in~\cite{noi}.

The ratios of the mean numbers of \kl\ per event to the charged multiplicity
drop in the first few bins (0 \les\ \Nch\ \les\ 6) and are roughly 
constant for \Nch\ \gt\ 6
(MB sample) for both \kz\ and \lz.
The dependence on \Nch\ is more pronounced for \kz\ than for \lz.
The fraction of \lz\ per event and per track is obviously smaller than that 
of \kz\ and for both is larger at 630 GeV than at 1800 GeV.
Finally, the dependencies of the number of \lz\ 
for the $soft$ and the MB samples on \Nch, besides differing by about a 
factor of two, 
are both roughly flat and different in shape from the corresponding 
of the \kz\ distributions.

\section{CONCLUSIONS}
\label{concl}

The present measurements extend the studies of charged particle properties 
in MB \pbarp\ interactions to \kz\ and \lz\ production.
Using the data available at the two c.m.s. energies obtained under the
same experimental conditions and similar statistics, we are able to
directly compare the \vv\ production properties at the two c.m.s. energies.
Our results offer new findings and significant improvements to the
existing knowledge of \vv\ production. 
We summarize our results as follows:
\begin{itemize}
\item The overall production rates of \kz\ and \lz\ are in agreement with 
      previous measurements.
\item The inclusive \pt\ spectra of \kz\ and \lz\ now extend to \pt\ \gt\ 
      8 \gevc. The \kz\ distribution shows a more detailed shape in 
      the high \pt\ region when compared to 
      previous data. For both \kz\ and \lz, we measure an average \pt\ 
      significantly higher than previous results.
\item New results  are presented on the distribution  of  
      \kz\ and \lz\ multiplicity.
\item For the first time, the MB sample has been used to analyze \vv\ 
      production properties in its $soft$ and $hard$
      components. Inclusive \pt\ and multiplicity distributions of \vv\
      are shown for the $soft$ and $hard$ data.
\item Analyses of the dependence of the mean \kl\ \pt\ with the event charged 
      multiplicity are presented.
      The observed dependence is not explained by the current theoretical
      models.
      Comparison with an analogous study performed on charged tracks indicates
      that the rate of the dependence grows with particle mass.
      An increase of the mean \pt\ is observable also in the $soft$ subsample
      alone.
\item The event charged multiplicity has been adopted as the independent 
      variable to analyze the ratio of the mean number of \kl\
      per event to the number of primary charged particles.
      For both \kz\ and \lz\ this ratio rises toward very low multiplicity,
      remaining roughly constant for \Nch\ \ges\ 5.  
\end{itemize}

\section*{ACKNOWLEDGEMENTS}
\label{ack}

We thank the Fermilab staff and the technical staffs of the participating 
institutions for their vital contributions. This work was supported by the 
U.S. Department of Energy and National Science Foundation; the Italian Istituto
Nazionale di Fisica Nucleare; the Ministry of Education, Culture, Sports, 
Science and Technology of Japan; the Natural Sciences and Engineering Research 
Council of Canada; the National Science Council of the Republic of China; the 
Swiss National Science Foundation; the A.P. Sloan Foundation; the 
Bundesministerium fuer Bildung und Forschung, Germany; the Korean Science and 
Engineering Foundation and the Korean Research Foundation; the Particle Physics
and Astronomy Research Council and the Royal Society, UK; the Russian 
Foundation for Basic Research; the Comision Interministerial de Ciencia y 
Tecnologia, Spain; in part by the European Community's Human Potential 
Programme under contract HPRN-CT-2002-00292; and the Academy of Finland.

%

%
\clearpage
\newpage
%
\begin{table}
\caption{\label{tab1} Raw and corrected numbers of \kz\ and \lz\ 
 found in each data set. In the rightmost three columns the fraction 
 of \kl\ per event is shown. The uncertainties on all the corrected numbers 
 and fraction of \kl\ per event include the systematic uncertainty.
 The total number of MB events at $\sqrt{s}$=1800 GeV (630 GeV) is
  2,079,558(1,963,157).}
\begin{ruledtabular}
\begin{tabular}{llccccccccc}
\multicolumn{2}{l}{} & \multicolumn{3}{c}{RAW} & 
\multicolumn{3}{c}{CORRECTED} & 
\multicolumn{3}{c}{FRACTION of \kl/EVENT (\%)} \\ 
\multicolumn{2}{l}{} & MB & Soft & Hard & MB ($\times10^{3}$) & 
  Soft ($\times10^{3}$) & Hard ($\times10^{3}$)& MB & 
          Soft & Hard \\ 
\hline
1800 & \kz\ & 36642 & 6733 & 29909 & 180$\pm$50 & 34$\pm$10 & 150$\pm$40 & 
      8.8$\pm$2.6 & 3.5$\pm$1.0 & 13.3$\pm$4.0 \\
GeV & \lz\ & 7518 & 782 & 6736 & 90$\pm$30 & 9$\pm$3 & 80$\pm$20 &
       4.3$\pm$1.0 & 1.0$\pm$0.3 & 7.2$\pm$2.2 \\
 & & & & & & & & & & \\
630 & \kz\ & 32222 & 9835 & 22387 & 170$\pm$50 & 50$\pm$15 & 120$\pm$35 &
       8.6$\pm$2.6 & 4.5$\pm$1.4 & 14.3$\pm$4.3 \\
GeV & \lz\ & 5883 & 1098 & 4785 & 70$\pm$20 & 13$\pm$4 & 60$\pm$20 & 
       3.7$\pm$1.1 & 1.2$\pm$0.4 & 7.1$\pm$2.1 \\ 
\end{tabular}
\end{ruledtabular}
\end{table}
\begin{table}
\caption{\label{tab2} \kz/$\pi$ ratio in each data set. 
 Data computed in the full \pt\ range and for \pt\gt\ 0.4 \gevc\
 are shown. The ratios are evaluated integrating the \pt\ distributions 
 by extrapolating the fitted function down to \pt$_{min}$=0. (0.4) \gevc. 
 Here efficiency corrections and systematic uncertainties are included.}
\begin{ruledtabular}
\begin{tabular}{ccccc}
       & \pt$_{min}$ (\gevc) & MB    & Soft  & Hard \\
\hline
1800   &  0.0 & 0.14$\pm$0.05 & 0.38$\pm$0.12 & 0.11$\pm$0.04 \\ 
 & 0.4 & 0.13 \mpp\ 0.04 & 0.30 \mpp\ 0.09 & 0.11 \mpp\ 0.03 \\
 & & & & \\
630    &  0.0 & 0.19$\pm$0.06 & 0.42$\pm$0.13 & 0.14$\pm$0.05 \\
 & 0.4 & 0.18 \mpp\ 0.05 & 0.33 \mpp\ 0.10 & 0.18 \mpp\ 0.06 \\
\end{tabular}
\end{ruledtabular}
\end{table}
\clearpage
\newpage
\begin{table}
\caption{\label{tab3} Summary of all systematic uncertainties. 
For each systematic uncertainty source its effect on the various measured 
quantities is reported. The symbol ``-'' means no effect on the 
corresponding quantity.}
\begin{ruledtabular}
\begin{tabular}{lcccccc}
Systematic & \ptm\ vs Mult. & \pt\ distr. & N(\vv) distr. & $<$N(\vv)$>$ vs Mult &
 N(\vv) & $K/\pi$ \\ 
uncertainty & (included & 
\multicolumn{2}{l}{(-------------------------} & not included in figures &
 ---------------) & (included \\
source & in figures) &  & & & & in table~\ref{tab2}) \\
\hline
MC simulation  & 3\% (\kz) & 10\%  & 10-25\% & 2-20\% & --- & --- \\
 & 4\% (\lz) &  &  &  &  &  \\
 & & & & & & \\
Low-\pt\   & 3-20 \%   & ---  & 5\% (\kz) & 5\% (\kz) & 5\% (\kz) & 6\% \\
extrapolation &  &  & 10\% (\lz) & 10\% (\lz) & 10\% (\lz) & \\ 
 & & & & & & \\ 
Primary vertex & --- & --- & 
\multicolumn{4}{c}{----------------------------------   $< 9\%$   
------------------------------} \\
selection & & & & & & \\
 & & & & & & \\
Global normalization & --- &
 \multicolumn{5}{c}{------------------------------------------   $<30\% $   
------------------------------------} \\
factor & & & & & & \\
\end{tabular}
\end{ruledtabular}
\end{table}
\begin{table}
\caption{\label{tab4} Number of events with $N_{K}$ \kz\ ($N_{\Lambda}$ \lz) 
divided by the total number
of events. MB, $soft$ and $hard$ data at 1800 and 630 GeV are reported.}
\begin{ruledtabular}
\begin{tabular}{lccccccc}
 & $N_{K/\Lambda}$ & 
\multicolumn{3}{c}{------------------- 1800 GeV -------------------} & 
\multicolumn{3}{c}{-------------------  630 GeV -------------------} \\ 
 &  & MB ($\times10^{-3}$) & Soft ($\times10^{-3}$) & Hard ($\times10^{-3}$) & 
   MB ($\times10^{-3}$) & Soft ($\times10^{-3}$) & Hard ($\times10^{-3}$) \\ 
\hline
 & 0 & 919\mpp\ 7 & 966\mpp\ 2 & 880\mpp\ 10 & 920\mpp\ 10 & 956\mpp\ 5 & 870\mpp\ 20 \\
 & 1 & 80\mpp\ 10 & 33\mpp\ 6& 110\mpp\ 20 & 80\mpp\ 20 & 40\mpp\ 10 & 120\mpp\ 30 \\
\kz\ & 2 & 5\mpp\ 1 & 0.9\mpp\ 0.2 & 9\mpp\ 2 & 4\mpp\ 1 & 1.2\mpp\ 0.3 & 8\mpp\ 2 \\
 & 3 & 0.6\mpp\ 0.2 & & 1.0\mpp\ 0.3 & 0.3\mpp\ 0.1 & 0.04\mpp\ 0.02 & 0.7\mpp\ 0.3 \\
 & 4 & 0.2\mpp\ 0.1 & & 0.4\mpp\ 0.2 & & & \\
\multicolumn{7}{c}{} \\
 & 0 & 958\mpp\ 5 & 990.3\mpp\ 0.1 & 930\mpp\ 10 & 964\mpp\ 5 & 988.3\mpp\ 0.1 & 930\mpp\ 10 \\
\lz\ & 1 & 40\mpp\ 20 & 10\mpp\ 4 & 70\mpp\ 30 & 40\mpp\ 20 & 12\mpp\ 5 & 70\mpp\ 30 \\
 & 2 & 1.6\mpp\ 0.8 & 0.08\mpp\ 0.05 & 3\mpp\ 1 & 0.9\mpp\ 0.5 & & 2\mpp\ 1 \\
\end{tabular}
\end{ruledtabular}
\end{table}
\clearpage
\newpage
\begin{table}
\caption{\label{tab5} Results of the fit to the invariant \pt\
 distribution of \kz. Data at different \ecms\ are reported (for 
 different experiments the parameters of the fit were 
 reported when available).
 Parameters $p^0$ and $n$ refer to the power-law (P.L.) function
 [Eq.(\ref{eq:fpt2})], $B$ to the exponential (Exp.) form [Eq.(\ref{eq:fpt5})].
 CDF-0 refers to the so called Run-0 of the Tevatron~\cite{CDFO} and 
 CDF-I to Run I data (this analysis).} 
\begin{ruledtabular}
\begin{tabular}{lcccccc}
Experiment & Data Set & $\overline{p_T}$ & $p^0$ (P.L.) & $n$ (P.L.) & B (Exp.) & \kdof\ \\
($\sqrt{s}$ in GeV) & & (\gevc) & (\gevc) & & & \\ 
\hline
UA5(*) (546)\cite{UA51} & MB & 0.58$\pm$0.04 & -- & -- & & 1.15  \\
CDF-0 (630)\cite{CDFO} & MB & 0.5$\pm$0.1 & 1.3(fixed) & 7.9$\pm$0.03& & 3.9 \\
CDF-I (630) & MB & 0.70$\pm$0.08 & 3.3$\pm$0.2 &
12.6$\pm$0.6 &  & 68/57 \\
UA5(*) (900)\cite{UA52} & MB & 0.63$\pm$0.03 & -- & -- & & 0.5  \\
CDF-0 (1800)\cite{CDFO} & MB & 0.60$\pm$0.03 & 1.3(fixed) & 7.7$\pm$0.2 & & 0.74 \\
CDF-I (1800) & MB & 0.58$\pm$0.02 & 1.3(fixed) & 7.49$\pm$0.02 & & 265/68 \\
CDF-I (1800) & MB & 0.75$\pm$0.07 & 3.29$\pm$0.08 &
11.7$\pm$0.1 &  & 67/67 \\
CDF-I (630) & Soft & 0.58$\pm$0.04 & 9.0$\pm$0.1 &
33.7$\pm$0.1 & & 29/22  \\
CDF-I (630) & Soft & 0.64$\pm$0.02 &  & & -3.12$\pm$0.03 & 24/23 \\
CDF-I (1800) & Soft & 0.62$\pm$0.02 & 9.5$\pm$0.3 &
33.7$\pm$0.9 & & 23/25 \\
CDF-I (1800) & Soft & 0.67$\pm$0.02 & & &-3.00$\pm$0.04 & 29/26 \\ 
\hline
\multicolumn{7}{l}{
(*) UA5 fits to a power-law form in $p_T>$0.4 together with
an exponential form in $p_T<$0.4~\gevc. } \\
\end{tabular}
\end{ruledtabular}
\end{table}
\clearpage
\newpage
\begin{table}
\caption{\label{tab6} Results of the fit to the invariant \pt\ 
 distribution of \lz. Data at different \ecms\ are reported (for 
 different experiments the parameters of the fit were reported 
 when available). Parameters $p^0$ and $n$ refer to the power-law 
 (P.L.) function [Eq.(\ref{eq:fpt2})], $B$ to the exponential (Exp.) 
 form [Eq.(\ref{eq:fpt5})]. CDF-I refers to Run I data (this analysis).}
\begin{ruledtabular}
\begin{tabular}{lcccccc}
Experiment & Data Set & $\overline{p_T}$ & $p^0$ (P.L.) & $n$ (P.L.) &
           B (Exp.) & \kdof\ \\
($\sqrt{s}$ in GeV) & & (\gevc) & (\gevc) & & & \\ 
\hline
UA5 (546)\cite{UA51} & MB & 0.62$\pm$0.08 & -- & -- & & --  \\
CDF-I (630) & MB & 0.91$\pm$0.07 & 12.3$\pm$0.1 & 30.1$\pm$0.2 & & 59/35 \\
CDF-I (630) & MB & 0.98$\pm$0.01 &  & & -2.05$\pm$0.03 & 50/36 \\
UA5 (900)\cite{UA52} & MB & 0.97$\pm$0.01 &  &  & -- & -- \\ 
CDF-I (1800) & MB & 0.97$\pm$0.09 & 12.4$\pm$0.1& 28.6 $\pm$ 0.09  & & 41/45 \\
CDF-I (1800) & MB & 1.04$\pm$0.01 & & & -1.92$\pm$0.02 & 55/46 \\
CDF-I (630) & Soft & 0.67$\pm$0.09 &  10.0$\pm$0.2 & 33.0$\pm$0.2 & & 31/24 \\
CDF-I (630) & Soft & 0.73$\pm$0.1 & & & -2.74$\pm$0.05 & 28/25 \\
CDF-I (1800) & Soft & 0.64$\pm$0.05 & 9.5$\pm$3.3 & 33.0 $\pm$0.2 & & 29/22  \\
CDF-I (1800) & Soft & 0.73$\pm$0.10 &  &  & -2.74$\pm$0.05 & 25/23 \\
\end{tabular}
\end{ruledtabular}
\end{table}
%
\clearpage
\newpage
%
\clearpage
\newpage
\clearpage
\newpage
\clearpage
\newpage
\clearpage
\newpage
\clearpage
\newpage
\clearpage
\newpage
\clearpage
\newpage
\clearpage
\newpage
\clearpage
\newpage
\clearpage
\newpage
\clearpage
\newpage
\clearpage
\newpage
\clearpage
\newpage
\clearpage
\newpage
\clearpage
\newpage
\clearpage
\newpage
\clearpage
\newpage
\clearpage
\newpage
\clearpage
\newpage
%

\begin{thebibliography}{99}

\bibitem{ua2} R. Ansari {\em et al.}, Z. Phys. C{\bf 36}, 175 (1987); 
              X. Wang and R. C. Hwa, Phys. Rev. D{\bf 39}, 187 (1989);
              UA1 Collaboration, F. Ceradini, Bari Europhys. High Energy 
              1985:0363. 
\bibitem{sjo} T. Sj\"{o}strand and M. van Zijl, Phys. Rev. D{\bf 36}, 2019 
              (1987), and references therein.
\bibitem{mod} A summary of the references to these models can be found 
              in~\cite{sjo}.
\bibitem{noi} D. Acosta {\em et al.}, Phys. Rev. D{\bf 65}, 072005 (2002);
\bibitem{detec} F. Abe {\em et al.}, Nucl. Instrum. Methods A{\bf 271}, 387 
                (1988), and references therein.
\bibitem{magnet} F. Abe {\em et al.}, Phys. Rev. D{\bf 52}, 4784 (1995). 
\bibitem{ctc} F. Bedeschi {\em et al.}, Nucl. Instrum. Methods A{\bf 268}, 50
                (1988). 
\bibitem{vtx} F. Snider {\em et al.}, Nucl. Instrum. Methods A{\bf 268}, 75 
              (1988).
\bibitem{calor} F. Balka {\em et al.}, Nucl. Instrum. Methods A{\bf 267}, 272
               (1988); S. R. Hahn {\em et al., ibid.} {\bf 267}, 351 (1988);
               K. Yasuoka {\em et al., ibid.} {\bf 267}, 315 (1988);
               R. G. Wagner {\em et al., ibid.} {\bf 267}, 330 (1988);
               T. Devlin {\em et al., ibid.} {\bf 268}, 24 (1988);
               S. Bertolucci {\em et al., ibid.} {\bf 267}, 301 (1988);
               Y. Fukui {\em et al., ibid.} {\bf 267}, 280 (1988);
               S. Cihangir {\em et al., ibid.} {\bf 267}, 249 (1988);
               G. Brandenburg {\em et al., ibid.} {\bf 267}, 257 (1988).
\bibitem{llbar} As we do not distinguish \lz\ from $\bar{\Lambda}^{0}$,
              in this paper \lz\ stands for both.
\bibitem{pdg} Particle Data Group, Phys. Rev D{\bf 66}, 010001-35 (2002);
              Phys. Rev D{\bf 66}, 010001-61 (2002).
\bibitem{multiparton} F. Abe {\em et al.}, Phys. Rev. D{\bf 56}, 3811 (1997);
              Phys. Rev. Lett. {\bf 79}, 584 (1997);\\
              X. Wang, Phys. Rev. D{\bf 46}, 1900 (1992);\\
              UA1 Collaboration, C. Albajar {\em et al}, Nucl. Phys. 
              B{\bf 309}, 405 (1988).
\bibitem{pythia} T.~Sj\"{o}strand, Computer Phys. Commun. {\bf 82}, 74 (1994);
                 G.Marchesini, B.R.Webber, G.Abbiendi, I.G.Knowles, 
                 M.H.Seymour,and L.Stanco, 
                 Computer Phys. Commun. {\bf 67},465 (1992).
\bibitem{ptfit} G. Arnison {\em et al.}, Phys. Lett. B{\bf 118}, 167 (1982);
              F. Abe {\em et al.}, Phys. Rev. Lett. {\bf 61}, 1819 (1988).
\bibitem{vanH} L. Van Hove, Phys. Lett. B{\bf 118}, 138 (1982);
               R. Hagedorn, Rev. Nuovo Cimento {\bf 6}, 10 (1983);
               J.D. Bjorken, Phys. Rev. D{\bf 27}, 140 (1983).
\bibitem{hwa} X. Wang and C. Hwa, Phys. Rev. D{\bf 39}, 187 (1989);
              M. Jacob, CERN preprint CERN/TH. 3515 (1983); 
              F.W. Bopp, P. Aurenche, and J. Ranft, Phys. Rev. D{\bf 33}, 
              1867 (1986).
\bibitem{CDFO} F.Abe {\em et al.}, CDF Collaboration, Phys. Rev. D{\bf 40}, 
               3791 (1989).
\bibitem{UA51} K.Alpgard {\em et al.}, UA5 Collaboration, Phys. Lett.
               B{\bf 115}, 65 (1982); G.J.Alner {\em et al.}, UA5 
               Collaboration, Nucl. Phys. B{\bf 258},505 (1985).
\bibitem{UA52} R.E.Ansorge {\em et al.}, UA5 Collaboration, Phys. Lett. 
               B{\bf 199}, 311 (1987).
\bibitem{E735} S.Banerjee {\em et al.}, E735 Collaboration, Phys. Rev. Lett.
               {\bf 62}, 12 (1989).
\bibitem{alex} T.Alexopoulus, II Int.Conf. on Phys. and Astrophys. of 
               Quark-Gluon Plasma, Calcutta, January 1993. 
\bibitem{barsh} S. Barshay, Phys.Lett. B{\bf 127}, 129 (1983).

\end{thebibliography}
\end{document}